\documentstyle[aps,epsfig,multicol]{revtex}

\def\rt{\rightarrow}

\def\beq{\begin{equation}}
\def\eeq{\end{equation}}

\input{epsf.tex}

\def\figone#1#2#3{\begin{figure}[h!]
\centering \leavevmode
\epsfxsize=0.75\columnwidth \epsfbox{#1}
\caption{\small #2 \label{#3}}
\end{figure}
}

\begin{document}

\title{Fluctuation-Dominated Phase Ordering Driven by Stochastically Evolving
         Surfaces}
\author{Dibyendu Das $^{1,2}$, Mustansir Barma $^1$\cite{byjnc} 
and Satya N. Majumdar $^{1,3}$}
\address{$^1$ Department of Theoretical Physics, Tata Institute of Fundamental Research, Homi Bhabha Road, Mumbai 400 005, India \\
$^2$ Martin Fisher School of Physics, Brandeis University, Mailstop 057, Waltham, Massachusetts 02454-9110, USA \\
$^3$ Laboratoire de Physique Quantique (UMR 5626 du CNRS), Universit\'e Paul Sabatier, 31062 Toulouse Cedex, France}
\maketitle

\begin{abstract}

We study a new kind of phase ordering phenomenon in
coarse-grained depth models of the hill-valley profile of fluctuating
surfaces with zero overall tilt, and for hard-core particles sliding
on such surfaces under gravity. We find that several such systems
approach an ordered state with large scale fluctuations which make
them qualitatively different from conventional phase ordered states.
We consider surfaces in the Edwards-Wilkinson
(EW), Kardar-Parisi-Zhang (KPZ) and Golubovic-Bruinsma-Das Sarma-Tamborenea
(GBDT) universality classes. For EW and KPZ surfaces,
coarse-grained depth models of the surface profile exhibit coarsening
to an ordered steady state in which the
order parameter has a broad distribution even in the thermodynamic limit,
 the distribution of particle
cluster sizes decays as a power-law (with an exponent $\theta$),
 and the 2-point spatial correlation function has a cusp (with
an exponent $\alpha = 1/2$) at small values of the argument.  The
latter feature indicates a deviation from the Porod law which holds
customarily, in coarsening with scalar order parameters. We present
several numerical and exact analytical results for the coarsening
process and the steady state. For linear surface models with dynamical
exponent $z$, we show that $\alpha = (z - 1)/2$ for $z < 3$, $\alpha =
1$ for $z > 3$, and there are logarithmic corrections for $z = 3$,
implying $\alpha = 1/2$ for the EW surface and $1$ for the GBDT
surface. Within the independent interval approximation we show that
$\alpha + \theta = 2$.  We also study the dynamics of hard-core
particles sliding locally downwards on these fluctuating
one-dimensional surfaces and find that the surface fluctuations lead to
large-scale clustering of the particles.  We find a
surface-fluctuation driven coarsening of initially randomly arranged
particles; the coarsening length scale grows as $\sim t^{1/z}$. The
scaled density-density correlation function of the sliding particles
shows a cusp with exponent $\alpha \simeq 0.5$, and $0.25$ for the EW
and KPZ surfaces. The particles on the GBDT surface show conventional
coarsening (Porod) behavior with $\alpha \simeq 1$.


\end{abstract}
\begin{multicols}{2}

\section{Introduction}

Phase ordering dynamics describes the way in which domains of an
ordered state develop when an initially disordered system is placed in
an environment which promotes ordering.  For instance, when a simple
ferromagnet or alloy is quenched rapidly from very high to very low
temperatures $T$, domains of equilibrium low-$T$ ordered phases form
and grow to macroscopic sizes.  A quantitative description of the
ordering process is provided by the time development of the two-point
correlation function; asymptotically, it is a function only of the
separation scaled by a length which increases with time, typically as
a power law \cite{Bray5}.

New phenomena and effects can arise when we deal with phase ordering
in systems which are approaching {\it nonequilibrium} steady states.  In
this paper, we study a coupled-field nonequilibrium system in which
one field evolves autonomously and influences the dynamics of the
other.  The system shows phase ordering of a new sort, whose principal
characteristic is that fluctuations are very strong and do not damp
down in the thermodynamic limit --- hence the term
fluctuation-dominated phase ordering (FDPO).

In usual phase ordering systems such as ferromagnetic Ising model, 
if one considers a finite system 
and waits for infinite time, then the system reaches
a state with magnetization per site very close to the two possible
values of the spontaneous magnetization, $m_s$ or $-m_s$,
with  very infrequent transitions between the two. This is reflected
in a probability distribution for the order parameter which is
sharply peaked at these two values, with the width of the peaks
approaching zero
in the thermodynamic limit (Fig. \ref{fdpo}(a)). By contrast, in the 
FDPO steady state, the system continually shows strong fluctuations in time
without, however, losing macroscopic order.  Accordingly, the order parameter
shows strong variations in time,
reflected eventually  in a probability distribution
which remains broad even in the thermodynamic
limit (Fig. \ref{fdpo}(b)).

The physical system we study consists of
an independently stochastically fluctuating surface of zero average slope, on which
reside particles which tend to slide downwards guided by the local
slopes of the surface.  Somewhat surprisingly, a
state with uniform particle density is unstable towards large scale
clustering under the action of surface fluctuations.
Eventually it is driven to a phase-ordered state with macroscopic
inhomogeneities of the density, of the FDPO sort.
Besides exhibiting a broad order parameter distribution, this
state shows unusual
scaling of two-point correlation functions and cluster
distributions.  It turns out that much of the physics
of this type of ordering is also captured by 
a simpler model involving a coarse-grained
characterization of the surface alone, and we study this as well.
A brief account of some of our results has appeared in \cite{das5}.

\figone{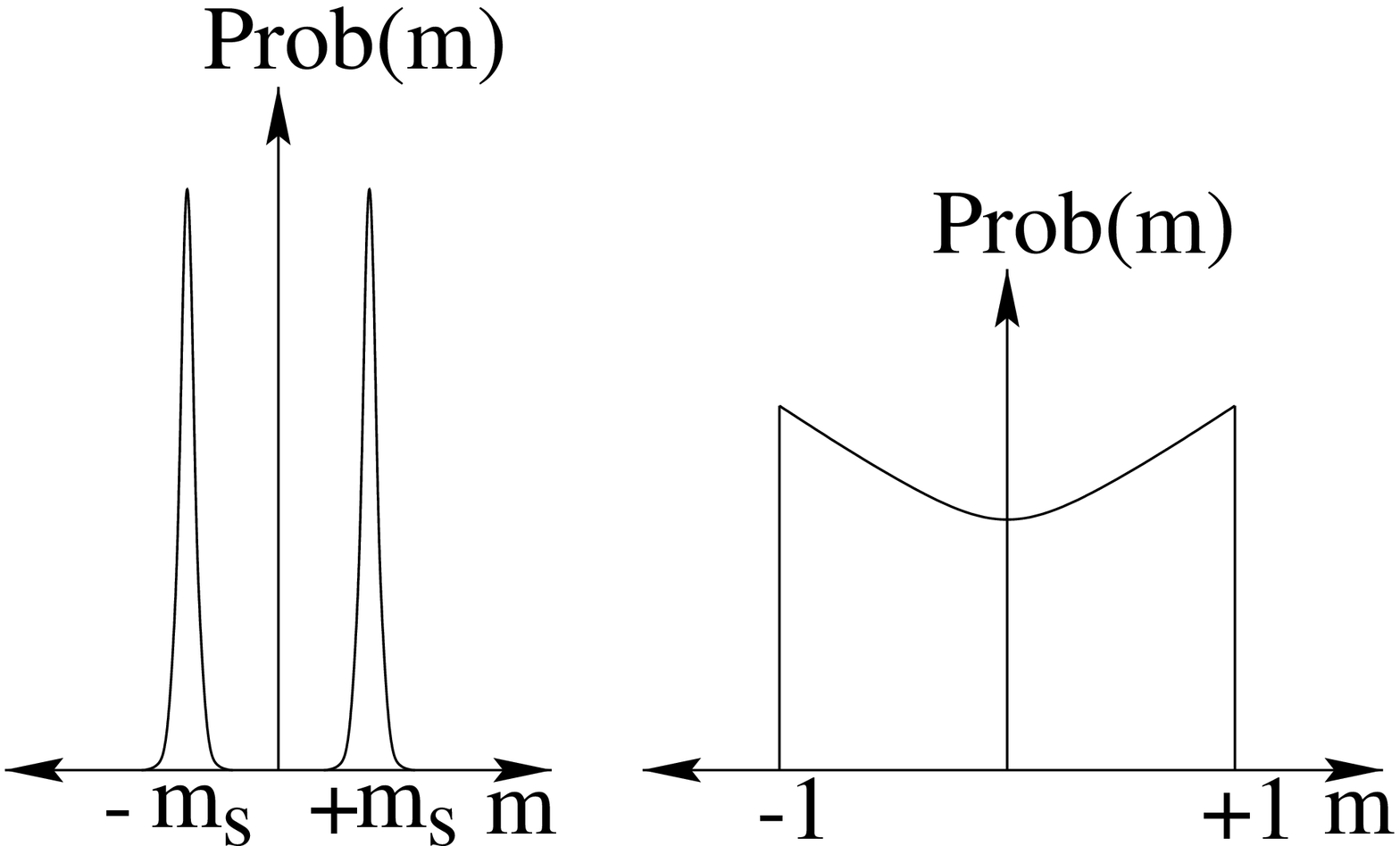}
{Schematic depiction of $Prob(m)$ against $m$ in steady state 
for $(a)$ a normal phase
ordering system such as a ferromagnet at low
temperature $(b)$ a system showing FDPO.}{fdpo}

In the remainder of the introduction, we first discuss the
characteristics of FDPO {\it vis a vis} normal phase-ordered states.
We then discuss, in a qualitative way, the occurrence of FDPO in
the surface-driven models under study. The layout of the rest of the
paper is as follows.  In Section \ref{FDPO_CD}, we define and study the
coarsening and steady states of three different coarse-grained depth
models of the fluctuating surfaces. In Section \ref{understandFDPO},
we demonstrate the existence of
a power-law in the cluster size distribution, and show
how it can give rise to FDPO.
In Section \ref{FDPO_SP}, we discuss ordering of sliding particles on
fluctuating surfaces.  In Section \ref{robust},
we explore the robustness of FDPO with respect to changes in various
rates defining the nonequilibrium process.
Finally, in Section \ref{Conclusion}
we summarize our principal results, and discuss the possible
occurrence of FDPO in models of other physical systems.

\subsection{Ordered States in Equilibrium Systems}

With the aim of bringing out the features of fluctuation dominated phase
ordering (FDPO) in nonequilibrium systems, let us recall some familiar
facts about phase ordered states in equilibrium statistical
systems.  We first discuss different characterizations of spontaneous
ordering, following the paper of Griffiths \cite{griffiths5} on the
magnetization of idealized ferromagnets. We follow this with a discussion
of fluctuations about the ordered state.

\subsubsection{Definitions of Spontaneous Order}

\noindent {\bf(a)} In the absence of a conservation law,
the magnetization $m$ is an indicator of the ordering:
\beq
m= { {1 \over
L^d} {\sum_{n}} s_n }
\label{defm11}
\eeq
where $L$ is the linear size, $d$ is the dimension and
$s_n$ is spin at site $n$.  In the thermodynamic limit,
the thermal average of the absolute value
\beq
m_1= {Lim_{L \rt \infty}} \langle
|m| \rangle~~({\rm nonconserved})
\label{defm1}
\eeq
with Boltzmann-Gibbs weights for configurations
provides an unequivocal measure of the order. This is because
in the low-temperature ordered phase,
the probability $Prob(m)$ of occurrence of  magnetization $m$ is
peaked at $+m_s$ and $-m_s$; the peak widths approach zero
in the thermodynamic limit $L \rt \infty$, so that
the average value $m_1$ coincides with the peak value $m_s$ (Fig.
\ref{fdpo}$(a)$).

For the conserved order parameter case, the value of the magnetization
is a constant and is same in both the disordered and ordered phase.
One therefore needs a quantity that is sensitive to the difference
between order and disorder. The simplest such quantity is the lowest
nonzero Fourier mode of the density \cite{Korniss5}
\beq
|Q| =  {1 \over L}{|{\sum_n} e^{2\pi i n/L}{(1+S_n)\over 2}|}
\label{defq*1}
\eeq
where $S_n$ denotes the average magnetization in the $(d-1)$-dimensional
plane $n$ oriented perpendicular to the $x$ direction.
The modulus in Eq. \ref{defq*1}  above
leads to the same value for all states which can be reached from each
other by a translational shift.
In the low-$T$ ordered phase, $Prob(Q)$ is expected to be a sharply peaked
function, with peak widths vanishing in the thermodynamic limit.
Then the mean value $Q_1$ defined by
\beq
Q_1= {Lim_{L \rt \infty}} \langle
|Q| \rangle~~({\rm conserved})
\label{defq*}
\eeq
serves as an order parameter.
A disordered state corresponds to $Q_1=0$, while a perfectly ordered state
with $m=1$ in half of the system and $m=-1$ in the other half
corresponds to $Q_1={1/\pi}\simeq 0.318.$

\noindent {\bf (b)} Another characterization of the order is obtained
 from the asymptotic
value of the 2-point spatial correlation function $C(r) = \langle
s_o s_{o+r} \rangle$. At large separations $r$, $C(r)$ is expected to
decouple:
\beq
{Lim_{r\rt\infty}} {Lim_{L\rt \infty}}\langle s_o s_{o+r}\rangle =
\langle s_o \rangle \langle s_{o+r} \rangle = m_{c}^{2} .
\label{defmc}
\eeq
A finite value of $m_c$ indicates that the system has long-range order.
A value $m_c = 1$ would indicate a perfectly
ordered pure phase without any droplets of the other species
(like the $T=0$ state of an Ising ferromagnet), while $m_c \neq 1$ would
indicate that the phase has an admixture of droplets of the other species
(like the state of an Ising ferromagnet for $0 < T < T_c$).

In a finite system, $C$ is a function only of the scaled
variable $r/L$ in the asymptotic scaling limit $r \rt \infty, L \rt \infty$
(see also {\bf (d)} below).
An operational way to find the value of $m_c$ is then
to read off the intercept ($r/L \rt 0$) in a plot of $C$ versus $r/L$; it gives
$m_{c}^{2}$ in the $L \rt \infty$ limit.

In equilibrium systems of the type discussed above, $m_1$
(defined in Eq. \ref{defm1}) and $m_c$ coincide.

\subsubsection{Characteristics of Fluctuations}

\noindent {\bf (c)} With a conserved scalar order parameter, the low-$T$ state
is phase-separated, with each phase occupying a macroscopically large region,
and separated from the other phase by an interface of width $W$.
The interfacial region is quite distinct from either phase, and on the
scale of system size, it is structureless and sharp.

\noindent {\bf (d)} Customarily in phase-ordered steady states, the spatial
correlation function $C(r)$ has a scaling form in $|r/L|$, for
$\xi << r << L$ where $L$ is the size of the system.
In the limit $r \rt \infty, L \rt \infty, |r/L| \rt 0$,
$C(r)$ follows the form \cite{Bray5}
\beq
C(r) \approx m_{c}^{2}(1 - 2|r/L|)~~~~(|r/L| \rt 0)
\label{porodr}
\eeq
The origin of the linear fall in Eq. \ref{porodr} is easy to understand
in systems  where phases are separated by sharp boundaries on the
scale of the system size, as in {\bf (c)} above: a spatial
averaging of $s_o s_{o+r}$ produces $+m_{c}^{2}$ with probability $(1-|r/L|)$
(within a phase) and $-m_{c}^{2}$ with probability $|r/L|$ (across phases).
 The linear drop with $|r/L|$ implies that the structure factor
$S(k)$, which is the Fourier transform of $C(r)$, is given, for large
wave-vectors $(kL>>1$),  by:
\beq
{S(k) \over L^d} \sim {1 \over {(kL)}^{d+1}}.
\label{porod}
\eeq
This form of the decay of the structure factor for scalar order parameters
is known as the Porod law.

It is worth remarking that the forms Eqs. \ref{porodr} and \ref{porod}
also describe the behaviour of the two-point correlation function in an
infinite system undergoing phase ordering starting from an initially
disordered  state. In such a case, $L$ denotes the coarsening time-dependent
length scale which is the characteristic size of an ordered domain.

\noindent {\bf (e)} For usual phase-ordered systems, spatial
fluctuations are negligible in the limit of the system size going to
infinity. Hence the averages of 1-point and 2-point functions over an
ensemble of configurations are well represented by a spatial average
for a single configuration in a large system.

\subsection{Fluctuation-Dominated Ordering}

The phase ordering of interest in this paper occurs in certain types
of nonequilibrium systems, and the resulting steady state
differs qualitatively from the ordered state
of  equilibrium systems and other types of nonequilibrium systems considered
earlier \cite{mukamel15}. The primary difference lies in the effects of fluctuations.
Customarily, fluctuations lead to large variations of the order parameter
which scale sublinearly with the volume, and so are negligible in
the thermodynamic limit.  Fluctuation effects are
much stronger here, and lead to variations of the order parameter in time,
without, however, losing the fact of ordering.
Below we discuss how the properties {\bf (a)-(e)} discussed above
are modified.

\noindent {\bf (a)} Nonzero values of the averages $M_1$ and $Q_1$
(Eqs. \ref{defm1} and \ref{defq*}) continue to indicate the existence of order,
but no longer provide an unequivocal measure of the order parameter.
This is because the probability distributions $Prob(m)$
and $Prob(Q)$ remain broad even in the limit $L \rt  \infty$ (as shown
schematically in Fig \ref{fdpo}$(b)$).

\noindent {\bf (b)} The measure $|m_c|$ of long-range order is nonzero,
and its value can be found from the intercept $C(|r/L| \rt 0)$.
However, the value of $m_c$ is, in general, quite different from $m_1$.

\noindent {\bf (c)}
As with usual ordered states, the
regions of pure phases are of the order of system size $L$. But in contrast
to the usual situation, there need not be a well-defined interfacial region,
distinct from either phase. Rather, the region between the two largest
phase stretches is typically a finite fraction of the system size, and
has a lot of structure; this region itself contains stretches of pure phases
separated by further such regions, and the pattern repeats.
Representative spin configurations $\{s_i\}$  for the two cases are
depicted schematically in Fig. \ref{cusp1}.  This nested
structure is consistent with a power-law distribution of cluster sizes,
and thus of a critical state. The crucial extra feature of the FDPO
state is that the largest clusters occupy a finite fraction of the total
volume, and it is this which leads to a finite value of $m_c$ as in (b)
above.
 Representative spin configurations $\{s_i\}$  for the two cases are
depicted schematically in Fig. \ref{cusp1}.

\figone{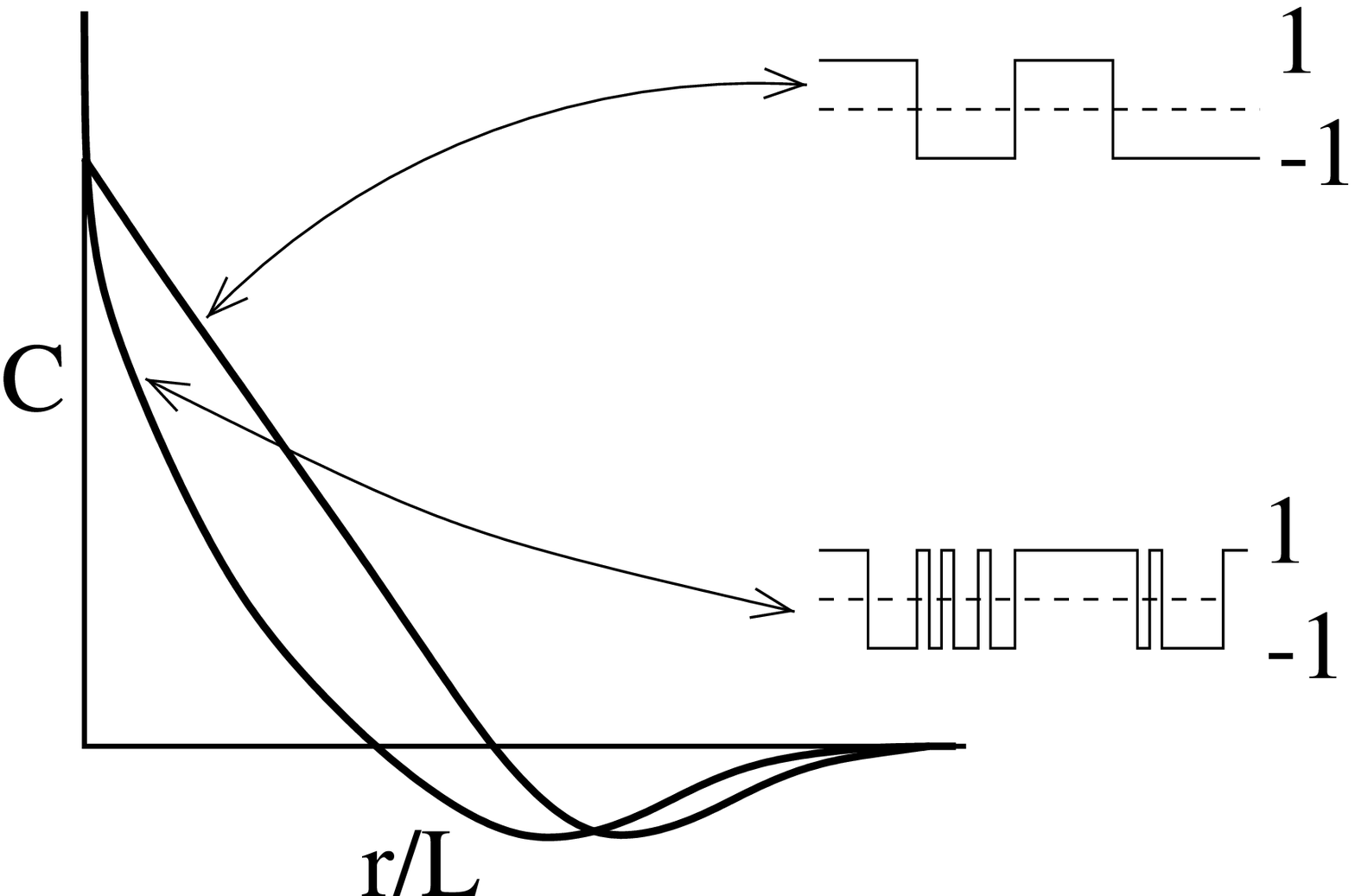}
{Schematic depiction of a linear and
a cuspy decay of $C(r)$ as a function of $r/L$, characteristic of normal phase
ordering and FDPO respectively. Typical configurations
corresponding to the two cases are also shown, with 1 and $-1$
denoting the values of $s_i$.}{cusp1}

\noindent {\bf (d)} The ensemble-averaged spatial correlation function $C(r)$
continues to show a scaling form in $|r/L|$. However, in contrast
to Eq. \ref{porod} it exhibits a cusp (Fig. \ref{cusp1}) at small values of
$|r/L|$:
\beq
C(r) \approx m_{c}^{2}(1 - b{|{r \over L}|}^{\alpha}).
\label{scale}
\eeq
This implies that the scaled structure factor varies as
\beq
{S(k) \over L^d} \sim {1 \over {(kL)}^{d+\alpha}}
\label{cusp}
\eeq
with $\alpha < 1$. This represents a marked deviation from the Porod law (Eq.
\ref{porod}). We will demonstrate in some cases that this deviation is
related to the power-law distribution of clusters in the interfacial region
separating the
domains of pure phases, as discussed in {\bf (c)} above.

\noindent {\bf (e)}
The spatial average of 1-point functions ($m$ or $Q$) and the
2-point function $C$ as a function of $|r/L|$ in a single configuration
of a large system typically
do not represent the answers obtained by
averaging over an ensemble of configurations. This reflects the
occurrence of macroscopic fluctuations.

\subsection{Fluctuating Surfaces and Sliding Particles}

Having described the general nature of fluctuation-dominated phase
ordering, we now
discuss the model systems that we have studied and which show 
FDPO. We consider
physical processes defined on a fluctuating surface with zero average
slope. The surface is assumed to have no overhangs, and so is characterized by a
single-valued local height variable $h(x,t)$ at position $x$ at time
$t$ as shown in Fig. \ref{sur1}. The evolution of the height profile
is taken to be governed by a stochastic equation.
The height-height correlation function
has a scaling form \cite{stanley5} for large separations of space and time:
\beq
\langle[h(x,t)-h(x',t')]^2\rangle \sim |x-x'|^{2\chi}f({|t-t'| \over
|x-x'|^z}).
\label{hhcorr}
\eeq
Here $f$ is a scaling function, and $\chi$ and $z$ are the roughness
and dynamical exponents, respectively. A common value of these
exponents and scaling function
for several different models of surface fluctuations indicate a
common universality class for such models. In this paper we will study
one-dimensional surfaces belonging to three such universality
classes of surface growth. Similar studies of two-dimensional surfaces
\cite{manoj} show that similar fluctuation-dominated  phase-ordered
states arise in these cases as well.

\figone{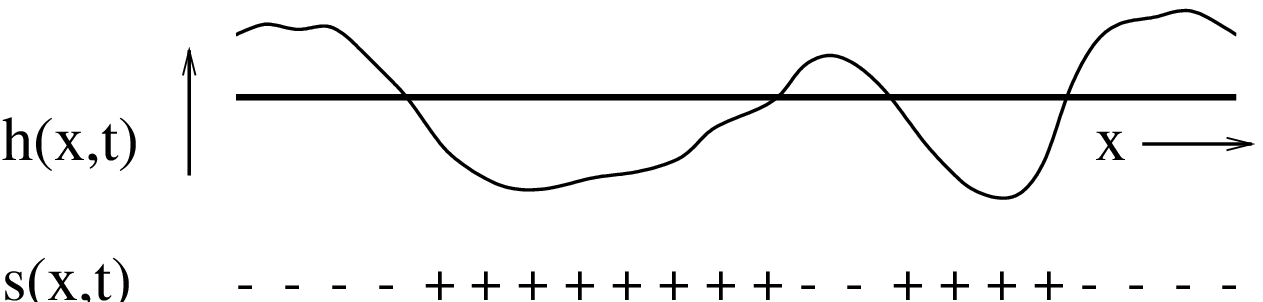}{Schematic depiction of a surface with height
$h(x,t)$ and the coarse-grained depth function $s(x,t)$.}{sur1}

Before turning to the physical model of particles sliding on such
fluctuating surfaces, we address the notion of phase ordering in
coarse-grained depth models associated with these surface
fluctuations. In Fig. \ref{sur1} we show the function $s(x,t)$ which
take values $+1$, $-1$ and $0$ depending on whether the height is
below, above or at the same level as some reference height $\langle h
\rangle$. Explicitly, we have $s(x,t) = -sgn(h(x,t) - \langle h
\rangle)$. Different definitions of $\langle h \rangle $ define
variants of the model; these are studied in Section II.

Starting from initially
flat surfaces, we study the coarsening of
of up-spin or down spin phases, which
arise from the evolution of surface profiles.  With the passage of time,
the surface gets rougher up to some length scale ${\cal L}(t)$. The
profile has hills and valleys, the base lengths of which
are of the order of ${\cal L}(t)$, implying
domains of like-valued $s$ whose size is of the same order.
Once the steady state is reached, there are landscape arrangements of the
order of the system size $L$ which occur on a time scale $L^z$. However,
these landscape fluctuations do not destroy long-range order, but
cause large fluctuations in its value.

Now let us turn to the problem of
hard-core particles sliding locally downwards under gravity
on these fluctuating surfaces.
Figure \ref{par1} depicts the evolution of particles
falling to the valley bottoms under gravity. When a local valley
forms in a region (Fig. 3(a) $\rightarrow$ Fig. 3(b)), particles in
that region tend to fall in and cluster together. The point is that particles
stay together even when there is a small reverse fluctuation
(valley $\rt$ hill as in Fig. \ref{par1}(b) $\rt$ (c)); declustering occurs
only if there is a rearrangement on length scales
larger than the size of the valley. The combination
of random surface fluctuations and the external force due to gravity
drive the system towards
large-scale clustering. Results of our numerical studies show that in
the coarsening regime, the typical scale of ordering in the
particle-hole system is comparable to the length scale over which
surface rearrangements take place. Further, the steady state of the particle
system exhibits uncommonly large fluctuations, reflecting the existence
of similar fluctuations in the underlying coarse-grained depth models
of the hill-valley profile. Similar effects are seen in 1-point
and 2-point correlation functions.
\figone{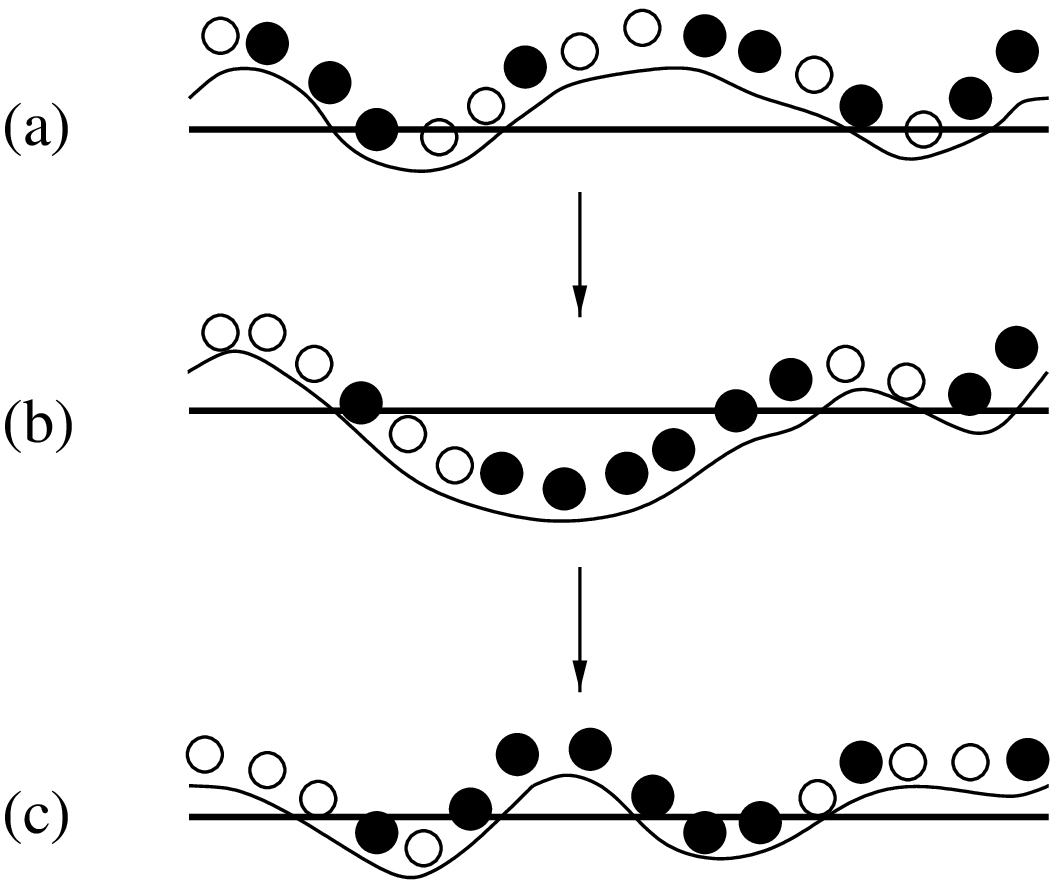}{Depicting clustering of
particles ($\bullet$) in a section of the fluctuating surface.
A surface fluctuation such as $(a) \rt (b)$
causes the particles to roll into a valley.
They remain clustered even after a local reverse surface fluctuation
$(b) \rt (c)$ occurs.}{par1}

\section{FDPO in coarse-grained depth (CD) models of surfaces}
\label{FDPO_CD}

\subsection{Surface evolution}
\label{surfevol}

The dynamics of surface fluctuations can be modelled by Langevin-type
equations for the height field $h(x,t)$. The evolution equations for
the one-dimensional Edwards-Wilkinson (EW)
\cite{Edwards5}, Kardar-Parisi-Zhang (KPZ) \cite{KPZ5}, and
Golubovic-Bruinsma-Das Sarma-Tamborenea (GBDT) \cite{DT5}
surface fluctuations are respectively
\begin{eqnarray}
EW:~{{\partial h} \over {\partial t}}& =& \nu_1 {{\partial^2 h} \over
{\partial x^2}} + \eta_1(x,t) \nonumber \\
KPZ:~{{\partial h} \over {\partial t}}& =& \nu_1 {{\partial^2 h} \over
{\partial x^2}} + \lambda ({{\partial h} \over {\partial x}})^2 +
\eta_1(x,t) \nonumber \\
GBDT:~{{\partial h} \over {\partial t}}& =& - K {{\partial^4 h} \over
{\partial x^4}} + \eta_1(x,t)
\label{surfaceeq}
\end{eqnarray}
where ~$\eta_1(x,t)$ ~is ~a ~white ~noise ~with
~$\langle \eta_1 \rangle = 0$ ~and
$\langle \eta_1(x',t') \eta_1(x,t) \rangle = \Gamma {\delta(x'-x)}
{\delta(t'-t)}$, and $\nu_1$, $\lambda$ and $K$ are constants.

In one dimension, the EW and KPZ models can be simulated using lattice
gas models whose large-distance large-time scaling properties coincide
with those of the corresponding continuum theories.
The lattice gas is composed of $\pm 1$-valued
variables $\{\tau_{i-{1 \over 2}}\}$ on a $1-d$ lattice with
periodic boundary conditions, where the ${\tau}$ spins occupy the links
between sites. The values $\tau_{i-{1 \over 2}} = +1$ or $-1$
represent the local slopes of the surface (denoted by $/$ or
$\backslash$, respectively). The dynamics of the interface is that of
the single-step model {\cite{Plischke5}}, with stochastic corner flips
involving exchange of adjacent $\tau$'s; thus, $/ \backslash
\rightarrow \backslash /$ with rate $p_{1}$, while $\backslash /
\rightarrow / \backslash$ with rate $q_{1}$.
For symmetric surface fluctuations ($p_{1} = q_{1}$),
the behavior at large length and time scale is described by the
continuum EW model. For $p_1 \neq q_1$, the surface evolution belongs to
the KPZ class. Corresponding to the configuration $\{\tau_{j-{1 \over 2}}\}$
we have the height profile $\{ h_i \}$
with $h_i = {\sum_{1{\leq}j{\leq}i}} \tau_{j-{1 \over 2}}$.

For simulating a surface fluctuating via a GBDT process, we used a
solid-on-solid model with depositing particles piling up
on top of each other. The height $h_i$ at site $i$ is the height of the pile
of particles at that site. During each micro-step
a particle is deposited randomly on a site $i$. If the new height $h_i$
at $i$, is greater than $h_{i-1}$ and $h_{i+1}$, then with equal
probability ($=1/3$) three things are attempted --- the deposited particle
can remain at site $i$, or can move to the neighboring sites
$i-1$ or $i+1$. It actually completes the
left or right move only if there is an increase in the coordination-ordination
number of the particles \cite{stanley5}.

\subsection{Definitions of the CD models}
\label{defcd}

Let us imagine a process of coarse-graining which eliminates fine fluctuations
of the height profile, and replaces the height field $h_i$ at site $i$
by a variable $s_i$ which is +1, -1 or 0 depending on whether the surface
profile at site $i$ is below, above or exactly coincident with a certain
reference level, which is the same at all $i$. The aim is to have a
coarse-grained construction of locations of large valleys and hills. Our
procedure depends on the choice of the reference level, and we have explored
three choices (the CD1, CD2 and CD3 models) which are discussed below.

In model CD1, the reference level is set by the initial condition,
which corresponds to an initially flat interface: $h(x,t=0)=0$. The
coarse-grained depth function is then
\beq
s(x,t) = -sgn[h(x,t)].
\label{cd1}
\eeq
With the passage of time, the surface becomes rougher, so that
$h(x,t)$ develops hills and valleys with respect to the $0$ level.
As the base lengths of the hills and valleys grow in size, there is a
growth of the domains of the variable $s(x,t)$. We are able to characterize the
coarsening behaviour of this model analytically in some cases.

In a finite system, at long enough times the surface  moves arbitrarily
far away from its initial location. Thus the steady state of the CD1 model
is trivial --- all $s_i$ are 1, or all are -1, with probability one.
This clearly happens because the reference level in the CD1 model is
fixed in space. This leads us to examine models (CD2 and CD3) where the
reference level moves along with the surface, so that we may expect
nontrivial steady state properties.

In model CD2, the coarse-grained depth function
\beq
s_{i} = - sgn[h_i]
\label{cd2}
\eeq
where $h_i = {\sum_{1{\leq}j{\leq}i}} \tau_{j-{1 \over 2}}$ as defined
in the earlier section. Note that at all times $t$, the origin is pinned
so that $h_{i=0} = 0$. The height function of the continuum version of
the CD2 model is related to that of CD1 as: $h_{j}^{CD2}(t) =
h_{j}^{CD1}(t) - h_{0}^{CD1}(t)$. The function $s_i$ is $+1$, $-1$ or $0$
accordingly as the height $h_i$ at site $i$ is below, above or at the zero
level. A stretch of like $s_i$'s $= +1$ represents a valley with respect
to the zero level. The time evolution of the CD2 model variables $\{ s_i \}$
is induced by the underlying dynamics of the bond variables
$\{ \tau_{i-{1 \over 2}} \}$ defined in the previous subsection.
This model was studied by us in \cite{das5}.

Finally model CD3, is defined as follows: $h_i$ is
constructed from $\tau$'s exactly as described
for the CD2 model, but then one defines
\beq
s_i = -sgn[h_i - \langle h(t) \rangle]
\label{cd3}
\eeq
where $\langle h(t) \rangle = (1/L)\sum_{i=1}^{L} h_i(t)$ is
the instantaneous average height which fluctuates with time.  This
definition was used earlier by Kim {\it et al} {\cite{Kim5}} who were
studying domain growth in an evolving KPZ surface.

Each of the CD models defined above has its own merits and
limitations.  We will see below that the CD1 model proves to be
analytically tractable (for Gaussian surface fluctuations) in
the coarsening regime, while for the CD2 model several exact results
can be derived in the steady state.  Of the three models, the CD3 model
most resembles the model of sliding hard-core particles on the surface
that is studied in Section \ref{FDPO_SP} below.

\subsection{Coarsening in the CD models}
\label{cdcoar}

\subsubsection{Analytical results for the CD1 model}

In this section, our primary focus is on coarsening properties of a
class of CD1 models. To this end, we will focus on the equal
time correlation function
\begin{equation}
C(x,t) = \langle\sigma(0,t)\sigma (x,t)\rangle =\langle sgn[h(0,t)]sgn[h(x,t)]\rangle .
\label{corr1}
\end{equation}
We consider
only linear interfaces evolving from a flat initial condition $h(x,0)=0$
according to the Langevin equation,
\begin{equation}
{{\partial h}\over {\partial t}}=-{\left ( -\nabla^2\right )}^{z/2} h +\eta,
\label{interface1}
\end{equation}
where $\eta (x,t)$ is a Gaussian white noise with $\langle \eta(x,t)\rangle=0$ and $\langle \eta
(x,t)\eta(x',t')\rangle=\delta(x-x')\delta(t-t')$. The dyanmic exponent $z$ specifies the relaxation mechanism.
For example, $z=2$ corresponds to Edwards-Wilkinson (EW) interface and $z=4$ corresponds to  Golubovic-Bruinsma-Das Sarma-Tamborenea (GBDT)
interface. Since $\eta(x,t)$ is a Gaussian noise and the evolution equation (\ref{interface1}) is linear, the
height field $h(x,t)$ is a Gaussian process. For Gaussian processes, it is straightforward
to evaluate the correlation function in Eq. (\ref{corr1}) exactly and one finds,
\begin{equation}
C(x,t) = {2\over {\pi}}{\sin}^{-1} \left[ H(x,t) \right],
\label{arcsine}
\end{equation}
where $H(x,t)$ is given by,
\begin{equation}
H(x,t)={ {\langle h(0,t)h(x,t)\rangle}\over { {\sqrt {\langle h^2(0,t)\rangle \langle
h^2(x,t)\rangle}
}}}.
\label{norm1}
\end{equation}
Now the normalized height correlation function $H(x,t)$
can be easily computed for linear interfaces evolving via Eq. (\ref{interface1}) by taking the Fourier transform of
Eq. (\ref{interface1}). From Eq. (\ref{interface1}), assuming flat initial condition, the Fourier transform,
$\langle h(k,t)h(-k,t)\rangle $ is given exactly by,
\begin{equation}
\langle h(k,t)h(-k,t)\rangle = {(1-e^{-2|k|^z t})\over {2|k|^z}}.
\label{ft1}
\end{equation}
Inverting this Fourier transform we get,
\begin{equation}
H(x,t)={ (z-1)\over { 2^{1-{1\over z}}\Gamma ({1\over {z}})} } F\left ({x\over {t^{1\over {z}}} }\right),
\label{scale1}
\end{equation}
where the scaling function $F(y)$ is given by,
\begin{equation}
F(y)=\int_0^{\infty} { {1-e^{-2u^z}}\over {u^z}} \cos (yu) du .
\label{scale2}
\end{equation}
Using this exact expression of $H(x,t)$ in Eq. (\ref{arcsine}), we get the exact correlation function
for arbitrary linear interface model parametrized by the dynamic exponent $z$. It is also evident that
$C(x,t)$ is a single function of the scaled distance, $y=xt^{-1/z}$.

The small distance behavior of the scaling function can be easily derived from the small argument asymptotics of
the integral in Eq. (\ref{scale2}). Let us first consider the EW interface with $z=2$.
In this case the integral in Eq. (\ref{scale2}) can be done (by putting a factor
$w$ in the exponential, i.e., writing $e^{-2wz^2}$ and then differentiating with respect to $w$ and then
integrating back with respect to $w$ upto $w=1$)
and we get,
\begin{equation}
H(x,t) = {1\over {2}}\int_{0}^{1} dw w^{-1/2} e^{-x^2/{8wt}}.
\label{e11}
\end{equation}
A change of variable, $x^2/{8wt}=y$ gives a more compact expression,
\begin{equation}
H(x,t) = {|x|\over {4\sqrt {2t}} }\int_{x^2/{8t}}^{\infty} e^{-y}y^{-3/2}dy.
\label{e12}
\end{equation}
Integration by parts yields the desired short distance behaviour,
\begin{equation}
H(x,t)= 1 - {\sqrt {\pi\over {8t}}}\,|x| + \ldots
\label{e13}
\end{equation}

Putting this back in Eq. (10) and expanding the arcsine, we get,
\begin{equation}
C(x,t) = 1 - {\left({2\over {\pi}}\right)}^{3/4} {| xt^{-1/2}|}^{1/2} + \ldots
\label{e14}
\end{equation}
Thus the correlation function has a square-root cusp at the origin for the $z=2$ CD1 model. One can similarly
do the small distance analysis for arbitrary $z>1$. We find that for general $z$,
\begin{equation}
C(x,t)=1- a |xt^{-1/z}|^{\alpha} +\ldots
\label{cuspform}
\end{equation}
where $a$ is a $z$-dependent constant and the cusp exponent $\alpha$ is given by,
\begin{eqnarray}
\alpha &= &(z-1)/2 \,\,\,\,\,  {\rm for} \,\,\,\,z<3 \cr
\alpha &=& 1  \,\,\,\,\,       {\rm for}\,\,\,\, z>3.
\label{e17}
\end{eqnarray}
For $z=3$, we find additional logarithmic corrections,
\begin{equation}
C(x,t)= 1 - a|y|{\sqrt {\log |y|}} + \ldots
\label{logz3}
\end{equation}
where $y=xt^{-1/3}$.

Thus our exact results indicate that $z=z_c=3$ is a critical value. For $z>3$, one recovers the
linear cusp in the correlation function at short distances (and hence Porod's law) indicating sharp interfaces between
domains
as in the usual phase ordering systems. But for $z<3$, one gets a $z$-dependent cusp exponent signalling
anomalous phase ordering dominated by strong fluctuations and a significant deviation from Porod's law. The value $z_c$ is the one across which a
morphological transition has been shown to occur in Gaussian surfaces
\cite{spa_per}, in the context of spatial persistence of fluctuating surfaces.

\subsubsection{Numerical results for the CD3 model}

Unlike the CD1 model, we have not been able to analytically characterize the coarsening properties
of the CD2 or CD3 models, in which the reference level moves with time. However
the coarsening properties in both CD2 and CD3 models can be studied numerically.
Below we present the numerical results for the equal time correlation function
${\cal C}$ for the CD3 model in three different cases where the underlying surface
is evolving respectively by the EW, KPZ and GBDT dynamics.
The initial condition chosen was $\tau_{j-{1\over2}} = 1$ at odd bonds
and $-1$ at even bond locations, ensuring that the height profile was globally
flat. We used a lattice with a number $L = 409600$ of bonds and equal number
of sites. At time $t > 0$ correlations gradually develop as the $s$-spin
domains grow. In Figs. \ref{ewcd3}, \ref{kpzcd3}, and \ref{dtcd3} we
show the data for ${\cal C}$ as a function of $t$
(insets of the respective figures), and how they collapse on to a single curve
${\cal C}_s$ in each case, on scaling $r$ by a $t$-dependent length scale
${\cal L}(t)$. For each of the three cases,
we see that ${\cal L}(t)
\sim t^{1/z}$, where the dynamical exponent $z = 2$, $3/2$, and $4$,
respectively for the EW, KPZ and GBDT surfaces. Notice that the
scaling curves for EW and KPZ surfaces have a cusp
at small values of the argument $r/{\cal L}$, and the cusp exponent
(Eq. \ref{scale}, \ref{cusp})
$\alpha \simeq 0.5$ for both. For the GBDT surface there is no cusp, and
$\alpha \simeq 1.0$. We note that these results for the CD3 model are consistent
with the analytical results in Eq. (\ref{e17}) of the CD1 model.

The fact that the correlation function has a scaling form in $r/{\cal
L}(t)$, with a nonzero intercept implies that at infinite time the
system would reach an ordered steady state, as the value of $\cal{C}$
at any fixed $r$ (no matter how large) approaches the value of the
intercept at large enough time. The intercepts of all the three curves
in Figs. \ref{ewcd3}, \ref{kpzcd3}, and \ref{dtcd3} have the value
$1$ implying that $m_c = 1$ for the CD3 model.

\figone{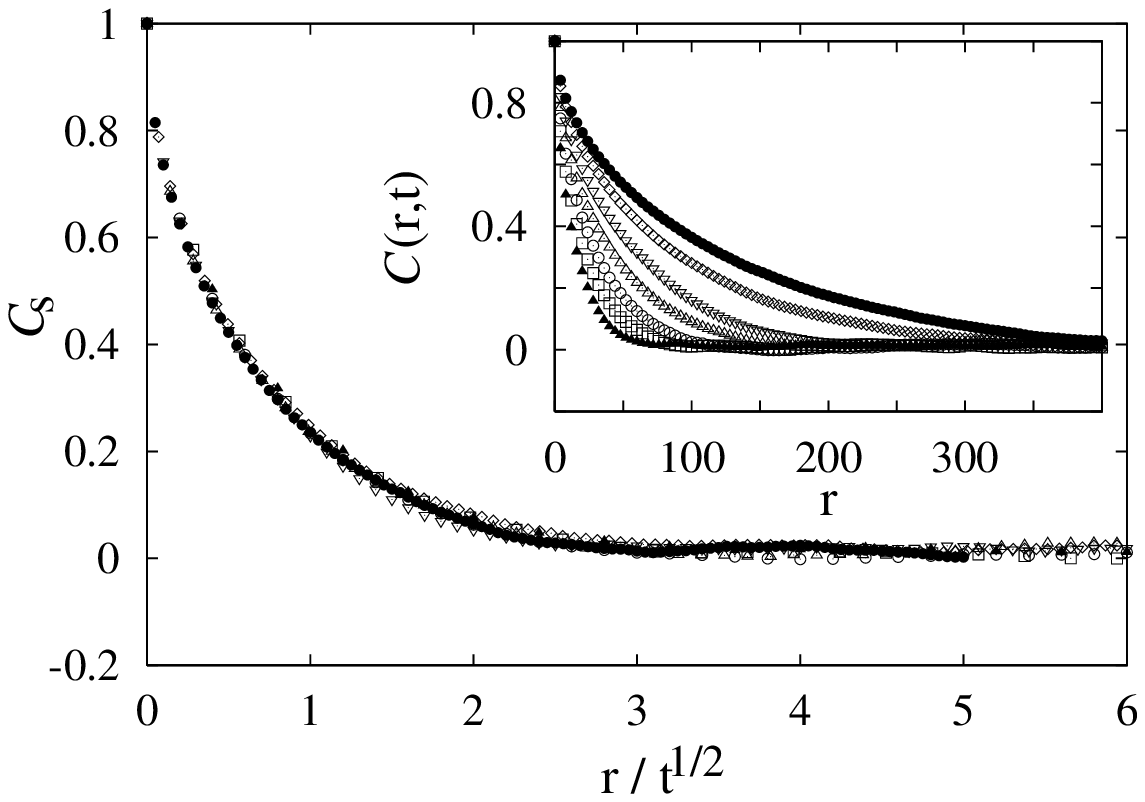}
{The data shown in the inset for ${\cal C}(r,t)$ for the CD3 
model of the EW surface 
at different times $t = 400{\times}2^n$ (with $n = 0$,...,$6$), 
is seen to collapse when $r$ is scaled by ${\cal L}(t) \sim t^{1/2}$.
The cusp in the scaling function at small argument is characterized by
 $\alpha \simeq 0.5$.}{ewcd3}

\figone{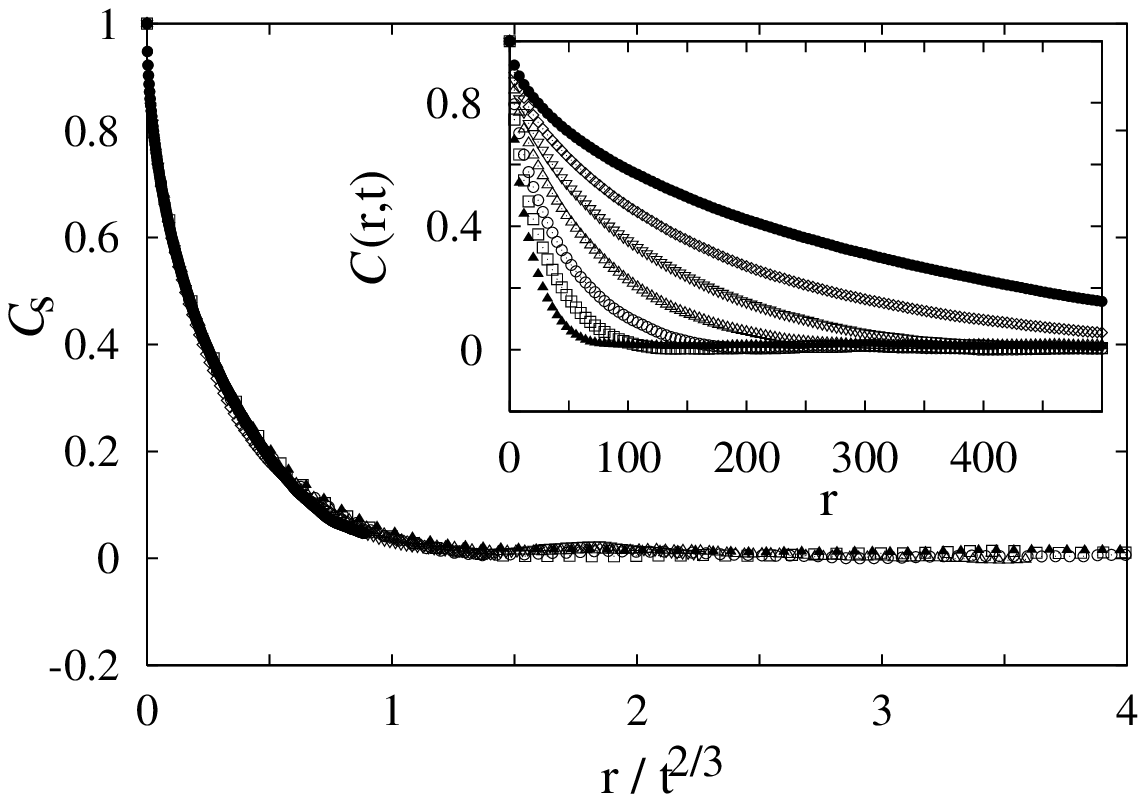}
{The data shown in the inset
for ${\cal C}(r,t)$ for the CD3 model of the KPZ surface
at different times $t = 400{\times}2^n$ (with $n = 0$,...,$6$),
is seen to collapse when $r$ is scaled by ${\cal L}(t) \sim t^{2/3}$.
The cusp in the scaling function at small argument is characterized by
 $\alpha \simeq 0.5$.}{kpzcd3}

\figone{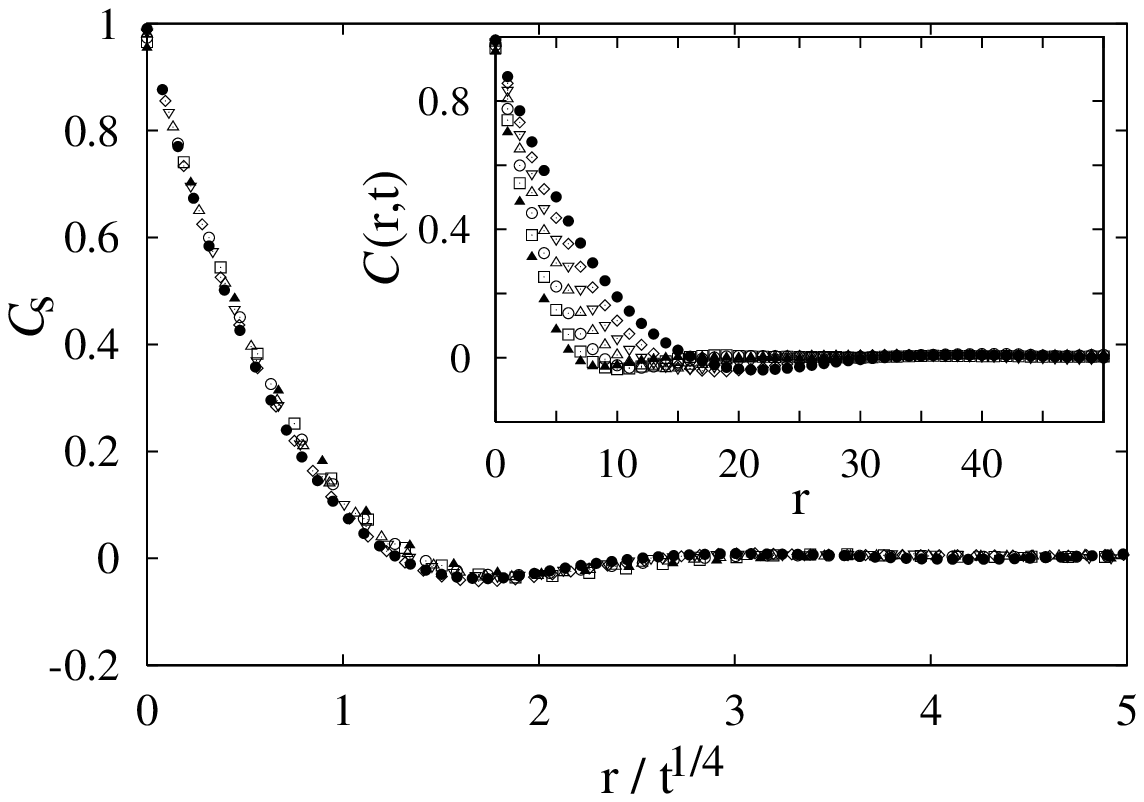}
{The data shown in the inset
for ${\cal C}(r,t)$ for the CD3 model of the GBDT surface
at different times $t = 400{\times}2^n$ (with $n = 0$,...,$6$), 
is seen to collapse when $r$ is scaled by ${\cal L}(t) \sim t^{1/4}$.
The behaviour of the scaling function at small argument is characterized by
$\alpha \simeq 1.0$.}{dtcd3}

\subsection{ Steady state of the CD models}
\label{cdst}

In a finite system, as time passes the surface diffuses away from its $t = 0$
location. As discussed above, this leads to a trivial steady state
in the CD1 models, corresponding to all $s_i=1$ (or all $s_i=-1$) with
probability one.
We need the reference level to keep up with the
surface in order to probe the steady state aspects of coarse-grained surface
fluctuations. This is accomplished in the CD2 and CD3 models.

In both the CD2 and CD3 models
we will see below that the cluster size
distribution of the $s_i$ variables varies as a power law $\sim
l^{-{\theta}}$ in the steady state. The order parameters have a broad distribution and the
scaled 2-point function has a cusp.

It is well known that for both EW and KPZ surfaces in $1-d$, the
steady states have random local slopes \cite{stanley5}, i.e. the steady
state probability distribution of the height profile is
\beq
P(\{h\}) = P_o e^{-\left[{{\int^{x'}}({{\partial h} \over
{\partial {x'}}})^2 dx'}\right]}.
\label{RW}
\eeq
This leads to a mapping of each surface configuration in the CD2 and
CD3 models to a random walk (RW) trajectory. The correspondence
is as follows: $\tau_{i-1/2}=+1$ or $-1$ can be interpreted as the
rightward or leftward RW step at the $i$'th time instant. Then
in the CD2 model, $s_i=1,-1$ or $0$ depending on whether the walker
is to the right, to the left or at the origin after the $i$'th step.
In the CD3 model, the reference point for demarcating left ($s_i = 1$)
and right ($s_i = -1$) is the average of displacements (heights), and
can be fixed only after the full trajectory is specified; then with
respect to $\langle h \rangle$,
the value of the position of the walker at every
$i$'th instant gets specified and hence also the $s_i$ spins.

\subsubsection{Power Law Distribution of Cluster Sizes}

For the CD2 model with EW or KPZ dynamics, exact results for different properties in the
steady state can be derived, because the surface profiles map on to
random walks. Periodic boundary conditions imply that the RW starts at
time $0$ from the origin and comes back to the origin after $L$ time
steps. Evidently, the lengths of clusters of $s=1$ spins (or $s=-1$
spins) represent times between successive returns to the origin. Thus
$P(l)$, the probability distribution of the cluster sizes $l$, for the
CD2 model is exactly the well-known distribution ($\approx 1/\sqrt{2\pi l^3}
e^{-{1/(2l)}}$) for RW return times
to the origin, which behaves as $\sim l^{-{3/2}}$ (for large $l$) with a
cutoff at $l = L$. Thus $\theta = 3/2$ in this model.

For the CD3 model, the variable reference point makes it difficult to make
exact statements, but we expect that the cluster
size distribution at large lengths $l$ will still be given as
$l^{-{3/2}}$. The numerically determined $P(l)$'s for the CD2 and CD3
models are plotted in Fig. \ref{plcd2cd3}, and they show the expected
power-law decay.

We note that the power-law distribution of the intervals between
successive returns is related to the spatial persistence of
fluctuating interfaces\cite{spa_per}. For linear interfaces
evolving via Eq. (\ref{interface1}), the corresponding
spatial persistence exponents were computed recently in
Ref. \cite{spa_per}. From these results we infer
that for CD models with GBDT dynamics, the cluster size distribution
in the steady state also has a power law distribution, $P(l)\sim l^{-5/4}$
in $(1+1)$-dimensions.

\figone{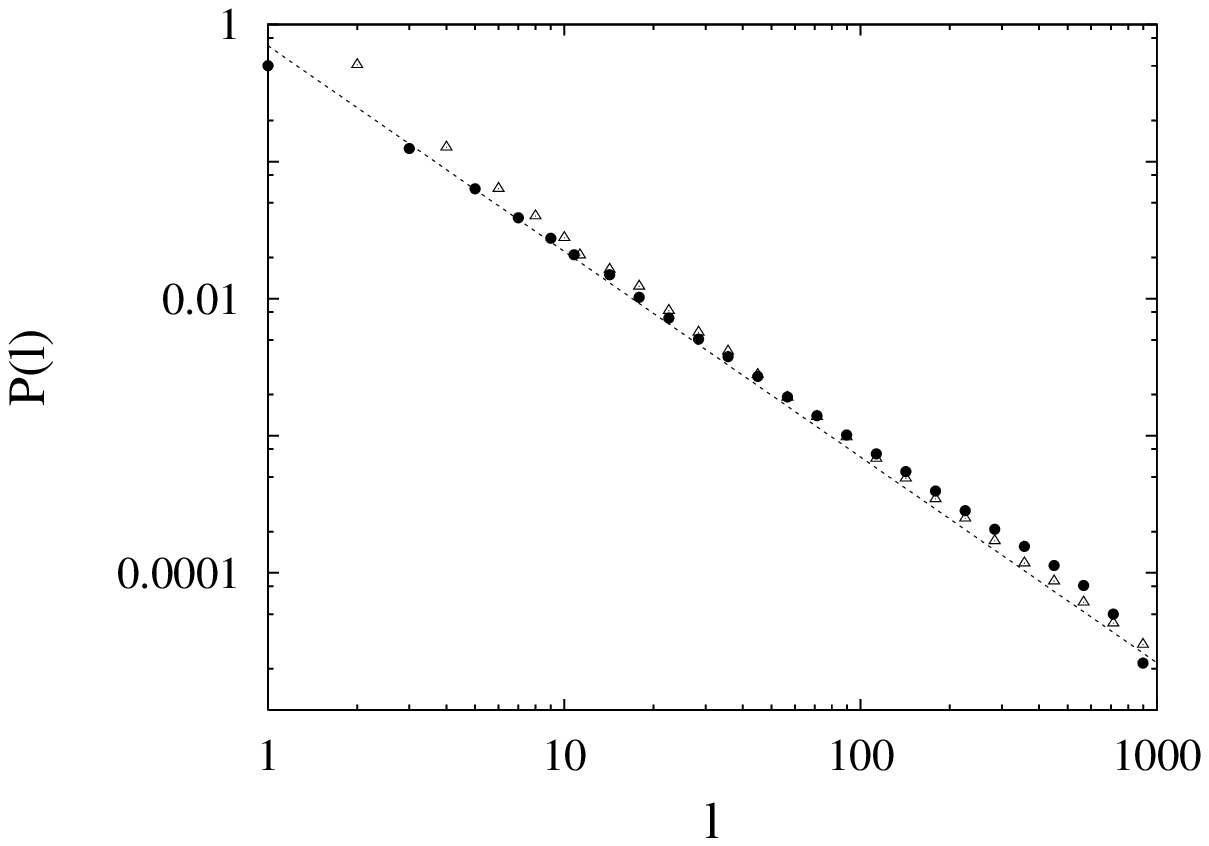}{$P(l)$ against $l$ for up-spin clusters
decays as $l^{-{3/2}}$ in both the
CD3 ($\bullet$) and CD2 (empty triangles) models.  We used $L=2048$.}{plcd2cd3}

\subsubsection{Order Parameter Distribution}

The distributions of the order parameters for each of the CD2 and CD3
models are broad. For the CD2 model, an appropriate (nonconserved)
order parameter is average value $m_1$ of modulus of
$m = {{1 \over L}\sum {s_i}}$ (see Eq. \ref{defm1}), which for the RW
represents the excess time a walker spends on one side of the origin
over the other side. In order to respect periodic boundary conditions,
we need to restrict the ensemble of RWs to those which return to the
origin after $L$ steps. The full probability distribution of $m$ over
this ensemble is known from the equidistribution theorem on sojourn
times of a RW \cite{Feller5}:
\beq
Prob(m) = 1/2~~~~~~~~,~~~~~~~~m \in [-1,1],
\label{probm1}
\eeq
i.e. every allowed value of $m$ is equally likely. This implies
${\langle |m| \rangle} = 1/2$ and $(\langle {m}^2 \rangle - {\langle |m|
\rangle}^2)^{1/2} = 1/{\sqrt 12}$.

For the CD3 model, most often half of the surface profile is above the
average height level and half below it. As a consequence, we find
numerically that the distribution of cluster sizes $P(l)$ decays sharply
beyond $L/2$. This resembles the sliding hard-core particles and hence
the conserved order parameter $Q_1$ is more suitable to
describe the ordering in this model than $m_1$. We monitor
the average value $Q_1$ of $Q^* = {1 \over L}{|{\sum_j}
e^{ij2\pi/L} \rho_j|}$ (Eq. \ref{defq*}), where $\rho_j =
(1+s_j)/2$. This order parameter has a value $0$ for a disordered
configuration and a value $1/\pi \approx 0.318$ for a fully phase
separated configuration with two domains of $+$ and $-$ spins, each of
length $L/2$. The numerical value of the distribution $P(Q^*)$ of
$Q^*$ is shown in Fig. \ref{probq1cd3}, and the average value
$Q_1$ in the limit of large system size numerically approaches
the value $0.22$. It is apparent from Fig. \ref{probq1cd3} that
$P(Q^*)$ is broad, and is larger for larger $Q^*$.  The width, which
remains finite in the thermodynamic limit, signifies that large-scale
fluctuations occur frequently in the system.

\figone{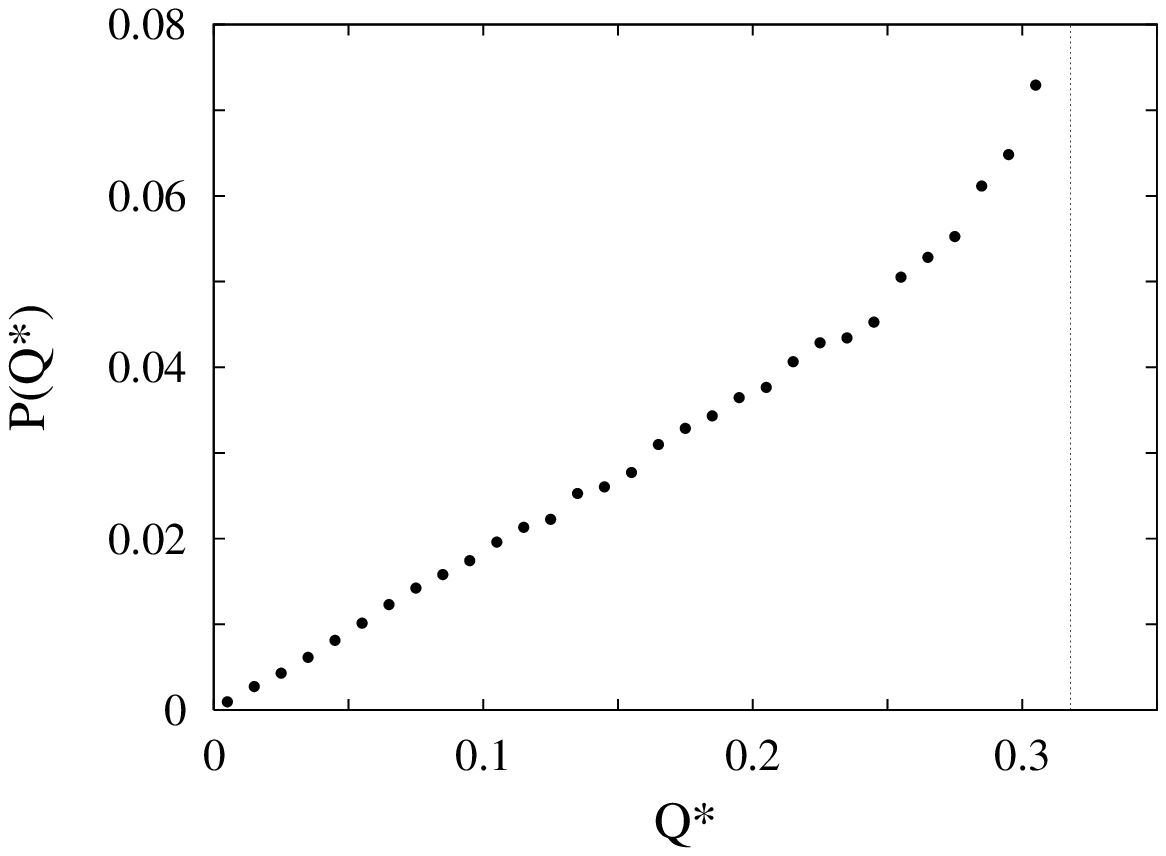}{Probability
distribution $P(Q^*)$ in the steady state of the CD3 model
The mean value is $Q_1 \simeq 0.22$.}{probq1cd3}

\subsubsection{Correlation Functions}

Finally we turn to the 2-point spatial correlation functions in the
steady state of CD2 and CD3 models. The growing length scale ${\cal L}(t)$
as $t \rt \infty$ is limited by the system size $L$.
In Fig. \ref{stcd2cd3} we show the
scaling of data for $C(r)$ in the steady state as a function of
$r/L$ for an EW surface. Both the curves show a cusp at
small values of $r/L$, with cusp exponent $\alpha \simeq 0.5$.

\figone{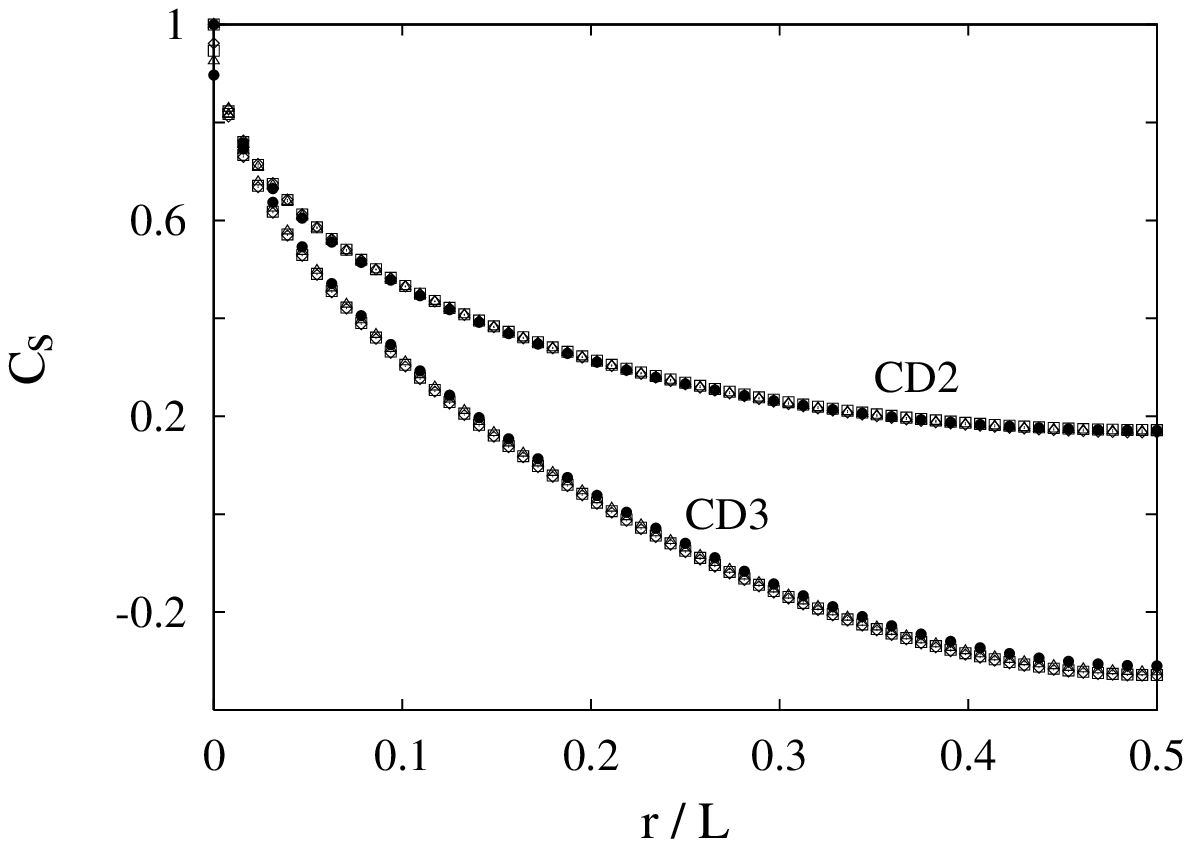}{Steady state
$C(r)$ collapses onto a single curves when plotted against $r/L$ for both the
CD2 and CD3 models.. The scaling
function shows a cusp at small values of the argument, with $\alpha \simeq
0.5$ for both models. We used $L = 64$, $128$, $256$ and $512$.}{stcd2cd3}

Since successive RW returns to the origin are independent events, the
calculation in section \ref{understandFDPO}A below based on independence of
intervals, is exact for the CD2 model. Thus Eq. \ref{iia3} holds, and we
conclude that the correlation function cusp exponent $\alpha =1/2$ exactly for
the CD2 model.  This also implies the result that $\alpha = 1/2$
even in the coarsening regime for the
CD2 model with EW and KPZ surfaces. This is because at any time $t$,
regions of a coarsening system which get equilibrated are of length $\sim
{\cal L}(t) << L$.  Now the correlation function ${\cal C}(r,t)$ is obtained
by spatial averaging over the system, and hence equivalently averaging
over an ensemble of several steady state configurations of subsystem
size $\sim {\cal L}(t)$. Thus the exact result for $\alpha$ in the
steady state carries over to the coarsening regime.

\section{Understanding FDPO in CD models}
\label{understandFDPO}

We have seen in the previous section that the distribution of like-spin
clusters follows a slow power law decay in the CD models.
We will demonstrate below that on the basis of this
power law, we may understand both (i) the cusp in the
2-point function, as well as (ii) ordered phases which occupy a finite
fraction of system size.

\subsection{Correlation Functions through the Independent Interval
Approximation}

We now show analytically within the Independent Interval Approximation (IIA)
{\cite{Satya5}} that the cusp exponent $\alpha$ and the power law exponent
$\theta$ are related. Within this scheme, the
joint probability of having $n$ successive intervals is treated as the
product of the distribution of single intervals. In our case, the intervals
are successive clusters of particles and holes, which occur with probability
$P(l)$. Defining the Laplace transform ${\tilde P}(s) = {\int_{0}^{\infty}}
dl e^{-ls} P(l)$, and ${\tilde C}(s)$ analogously, we have {\cite{Satya5}}
\begin{equation}
s(1 - s{\tilde C}(s)) = {2 \over {\langle l \rangle}}
{{1 - {\tilde P}(s)} \over {1 + {\tilde P}(s)}}
\label{iia1}
\end{equation}
where $\langle l \rangle$ is the mean cluster size. In usual applications of
the IIA, the interval distribution $P(l)$ has a finite first moment
${\langle l \rangle}$ independent of $L$. But that is not the case here, as
$P(l)$ decays as a slow power law $P(l) \sim l^{-{\theta}}\Theta(L-l)$ for
$l >> 1$, with the function $\Theta(L-l)$ denoting that the largest
possible value of $l$ is $L$. This implies that ${\langle l \rangle}
\approx a L^{2-\theta}$ for large enough $L$. Considering $s$ in the range
$1/L << s << 1$, we may expand ${\tilde P}(s) \approx 1 - b s^{\theta - 1}$;
then to leading order, the right hand side of Eq. ({\ref{iia1}}) becomes
$bs^{\theta - 1}/aL^{2-\theta}$, implying ${\tilde C}(s) \approx
1/s - b/(a L^{2-\theta} s^{3 - \theta})$. This leads to
\begin{equation}
C(r) \approx 1 - {b  \over {a \Gamma(3 - \theta)}}|{r \over L}|^{2 - \theta}.
\label{iia2}
\end{equation}
This has the same scaling form as Eq. ({\ref{scale}}).
Matching the cusp singularity in Eqs. (\ref{scale}) and (\ref{iia2}), we get
\begin{equation}
\theta + \alpha = 2~~~~~~~~~~~~~~~~~~~~~~~~~~~(\rm IIA).
\label{iia3}
\end{equation}
We recall (see Section \ref{cdst}) that the assumption of independent
intervals which underlies the IIA in fact holds exactly for the CD2 model,
and Eq. \ref{iia3} implies that $\alpha = 1/2$
in the steady state and the coarsening regime for the CD2 model.
For other models like the CD3 model, or the sliding particle
models we will encounter in the subsequent sections, the IIA gives
insight into the origin of the cusp from the power laws, although it is
not exact.

\subsection{Extremal Clusters and Ordered Phases}

We now turn to our claim (ii), that the very same distribution which
gives rise to power-law distributed broad boundaries with a collection
of small clusters, also gives rise to large clusters of size $\sim L$ of
`up' or `down' spins, which form the pure phases.
For the CD2 model, we numerically studied the sizes of the largest cluster
$l_1$ for systems of different sizes $L$; we show them in Fig.
\ref{largestL}. The full distribution ${\tilde P}(l_1)$ scales
as a function of $l_1/L$. The average
value is $\langle l_1 \rangle \simeq 0.48 L$. We also find a similar scaling
of the distribution ${\tilde P}(l_2)$, for the second largest clusters of
size $l_2$, and $\langle l_2 \rangle \simeq 0.16L$ (see Fig.
\ref{largestL}).

Some understanding of the fact that the size of largest clusters are
of order $L$ can be reached by considering the statistics of extreme
values. Applied to our case, if $N$ cluster lengths are drawn at
random from a distribution of lengths given by $P(l) \sim
(\theta-1)/l^{\theta}$, then the probability distribution $L_N (x)$
that the largest cluster is of length $x$ goes as $\approx N
x^{-\theta }exp(-Nx^{-(\theta -1)})$ \cite{zia5}.  The latter
distribution peaks at $x = x_{max} \sim N^{1/{(\theta -1)}}$. In the
CD2 problem $\theta = 3/2$. Now, in a system of length $L$ we
have on an average $\sqrt{L}$ clusters. If we make the approximate
replacement of $N$ by this average number $\sqrt{L}$, we immediately
get $x_{max} \sim L$. This explains how, although the average cluster
sizes are of order $L^{1/2}$, there are always clusters with sizes
of order $L$. This is reminiscent of the behaviour of the largest
loops in a random walk \cite{kantor}.

\figone{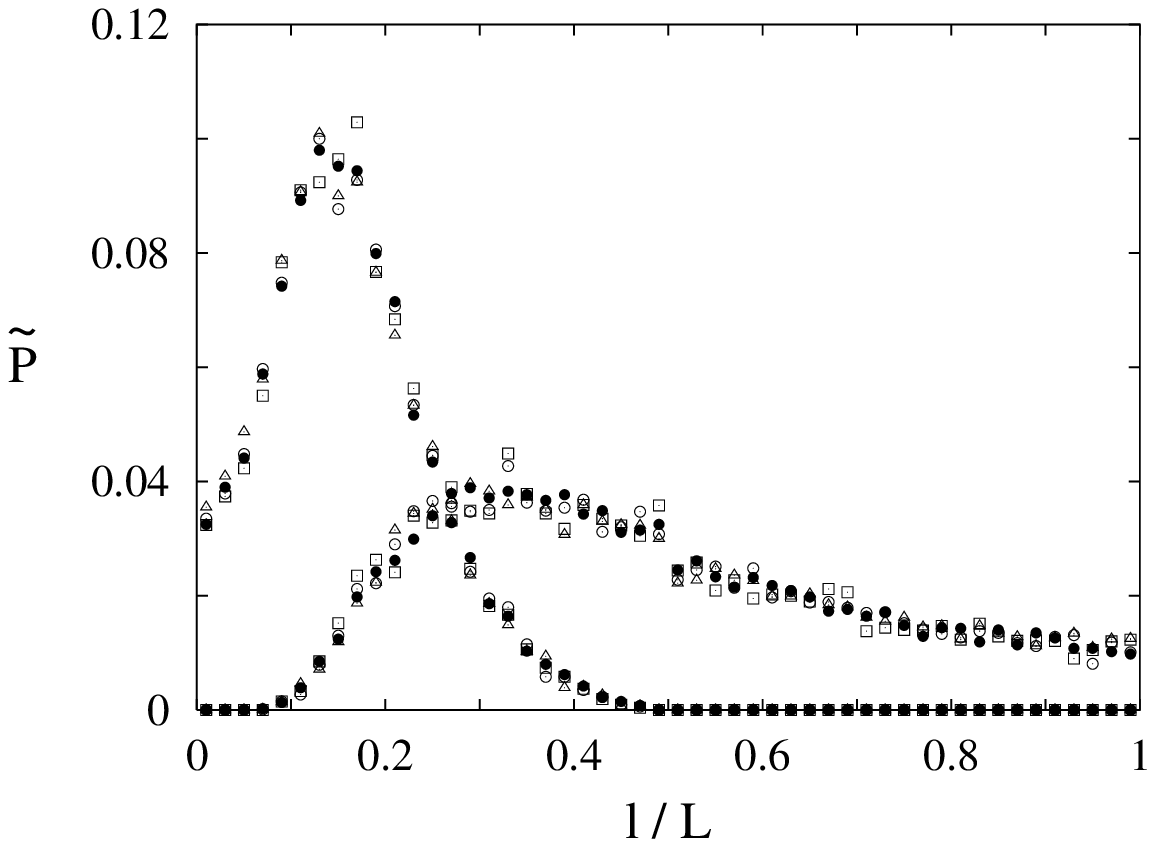}{Probability
distributions ${\tilde P}(l_1)$ (the curve on the right)
 and ${\tilde P}(l_2)$ of the
largest (length $l_1$) and second largest (length $l_2$) clusters in
the steady state of CD2 model are seen to collapse when plotted against
scaled lengths $l_1/L$ and $l_2/L$, respectively. The sizes used are
$L = 512$, $1024$, $2048$ and $4096$.}{largestL}

Further we found the contribution to magnetization coming from the
largest clusters in the system and compared them with the total
magnetization of the system, configuration by configuration. In
Fig. \ref{largemag}, we show scatter plots of
${\tilde m}_1$ which is the magnetization
obtained from summing the spins of the largest cluster, ${\tilde m}_2$
which is obtained by summing spins of largest and the second largest
cluster, and ${\tilde m}_3$ by summing those down to the third largest
cluster against the total magnetization $m = (1/L) \sum s_i$. The
convergence of the scatter plots towards the $45^o$ line, shows that
the few largest clusters give a major contribution to the
magnetization of the system.  Each of these large clusters is a pure
phase with magnetization $1$, and thus gives rise to $m_c=1$ in the
curves in Fig \ref{stcd2cd3}.

\figone{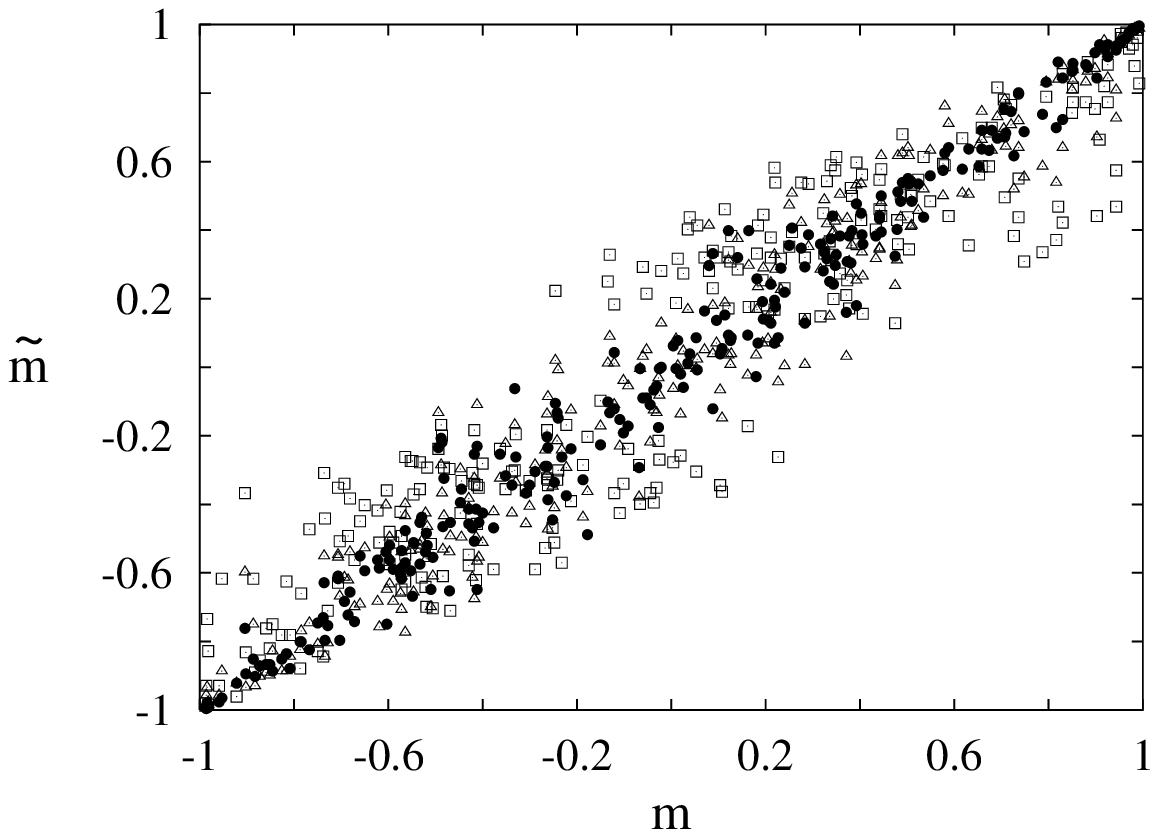}{ Estimates of the
magnetization ${\tilde m}_1$ (squares),
${\tilde m}_2$ (triangles) and ${\tilde m}_3$ ($\bullet$) from the largest
few clusters plotted against the total magnetization
for $m$, for $1000$ different configurations. The  convergence 
towards the line of slope unity shows 
that a few large clusters account for the major contribution to $m$.}
{largemag}

\section{Hard-core particles sliding on fluctuating surfaces}
\label{FDPO_SP}

\subsection{The Sliding Particle (SP) Model}

In this section we consider the physical problem of hard-core particles
sliding locally downwards on the fluctuating surfaces discussed in the
previous sections. We find that the downward gravitational force
combined with local surface fluctuations lead to large scale
clustering of the hard-core particles. The phase-separated state which
arises mirrors  the hill-valley profile
of the underlying surface.  For example, the
particles on EW and KPZ surfaces show FDPO with the 
cluster distribution, 1-point function,
and 2-point function behaving as in their CD model counterparts. On the other
hand, particles on the GBDT surface show conventional ordering.

Let us first define a sliding particle (SP) model on a one-dimensional
lattice. This is a lattice model whose behaviour resembles that
depicted in Fig. \ref{par1}.
The particles are represented by $\pm 1$-valued Ising
variables $\{\sigma_{i}\}$ on a one-dimensional lattice with periodic
boundary conditions, where $\sigma$ spins occupy lattice sites. The
$\tau_{i-{1\over 2}}$ variables occupy the bond locations and represent
the surface degrees of freedom as described in  section \ref{defcd}
for the CD2 model, and their dynamics involves independent evolution via
rates $p_1$ and $q_1$ as discussed earlier.  For the particles, ${\rho}_{i}
= {1 \over 2}(1 + {\sigma_{i}})$ represents the occupation of site
$i$. A particle and a hole on adjacent sites ($i$,$i+1$) exchange with
rates that depend on the intervening local slope $\tau_{i-{1 \over
2}}$; thus the moves $\bullet
\backslash \circ \rightarrow \circ \backslash \bullet$ and $\circ / \bullet
\rightarrow \bullet / \circ$ occur at rate $p_{2}$, while the inverse moves
occur with rate $q_{2} < p_{2}$. The asymmetry of the rates reflects the
fact that it is easier to move downwards along the
gravitational field. For most of our studies we consider
the strong-field ($q_2 = 0$) limit for the particle system. We set
$p_2 = p_1$. The dynamics conserves $\sum \sigma$ and $\sum
\tau$; we work in the sector where both vanish. This corresponds to
a $1/2$ filled system of particles on a surface with
zero average tilt. For the EW surface, we took
$p_1 = q_1$, while for the KPZ surface we took $p_1 = 1$ and $q_1 = 0$.
In Section \ref{robust}, we discuss departures from these conditions 
and explore the robustness of FDPO to these changes.

For the GBDT surface, the evolution of which was described in Section
\ref{surfevol}, a chosen particle moves to its right or left with
equal probability ($=1/2$) if there is locally a non-increasing height
gradient. Thus again $q_2 = 0$. The rate of update of the particles
is same as that of the surface.

The problem can be specified at a coarse-grained mesoscopic level
by the continuum equations for the density field
$\rho(x,t)$ corresponding to the discrete variable ${\rho}_i$ 
for the particles. Since the particle density is conserved,
the starting point is the continuity equation
${\partial \rho}/{\partial t} = - {\partial J(x,t)}/{\partial x}$, where
$J$ is the local current. Under the hydrodynamic assumption, the systematic
part of the above current is $-\rho {\partial h}/{\partial x}$, since for
viscous dynamics, the speed is proportional to the local
field, in this case the local gradient of height. Moreover there is
a diffusive part $- \nu_2 {\partial \rho}/{\partial x}$ which is driven
by local density inhomogeneities, and a noisy part $\eta_2(x,t)$ which
arises from the stochasticity. The noise $\eta_2$ is a Gaussian white noise.
The total density can be written as $\rho = \rho_o + {\tilde \rho}$, where
$\rho_o$ is the average density and $\tilde \rho$ is the
fluctuating part. This implies finally that the density fluctuation
$\tilde \rho$ evolves via the following equation:
\begin{eqnarray}
{{\partial {\tilde \rho}} \over {\partial t}} &=& \nu_2 {{\partial^2 {\tilde
\rho}} \over {\partial x^2}} + \rho_o (1-\rho_o)
{{\partial^2 h} \over {\partial x^2}} \nonumber \\
&+& (1-2\rho_o - 2{\tilde{\rho}})({{\partial {\tilde \rho}}
\over {\partial x}})({{\partial h} \over {\partial x}})
+ (1-2\rho_o){\tilde \rho}({{\partial^2 h} \over {\partial x^2}}) \nonumber \\
&-& {\tilde \rho}^2({{\partial^2 h} \over {\partial x^2}})
+ {{\partial \eta_2(x,t)} \over {\partial x}}
\label{particleeq}
\end{eqnarray}
Using the well-known mapping in $1-d$ between the density
field $\tilde \rho$ and the height field 
$\tilde h$ of the corresponding interface problem \cite{krugspohn}, 
one has the relation $\tilde \rho =
{\partial {\tilde h}}/{\partial x}$. This implies from Eq.
\ref{particleeq} that
the lowest order term in the evolution equation of $\tilde h$ is
proportional to ${\partial h}/{\partial x}$. This linear first-order
gradient
term is the result of the gravitational field which acts on the particles.
The evolution of the field $h(x,t)$ is given by Eq. \ref{surfaceeq}.
Thus a continuum approach to the problem of the sliding
particles requires analysis of the semi-autonomous set of nonlinear
equations \ref{surfaceeq} and \ref{particleeq} as  one of the fields
evolves independently but influences the evolution of the other. The problem 
belongs to the general class of semiautonomous systems, like the 
advection of a passive scalar in a fluid system \cite{kraichnan}.

The SP model is a special case of the Lahiri-Ramaswamy
(LR) model \cite{lahiri5,lahiri2} of driven lattices such as sedimenting
colloidal crystals. The general LR model has two-way linear couplings
between the $\rho$ and $h$ fields, and its phase diagram
has recently been discussed in \cite{4authors}. The SP model of interest
here has autonomous evolution of the $\{h(x)\}$, and corresponds
to the LR critical line which separates a wave-carrying phase \cite{waves}
from a strongly phase separated state \cite{lahiri2}.
Further, in a model of growing binary films considered in \cite{Drossel5}, in
the limit where the height profile evolves independently,
the problem gets mapped to noninteracting domain walls (if annihilation is
neglected) rolling down slopes of independently growing surfaces. The
latter problem becomes similar to ours, on thinking of the domain walls
as particles. But the fact that they are noninteracting in contrast to the
hard-core particles may introduce other physical effects into the problem.

\subsection{Coarsening in SP model}
\label{spcoar}

We start with a surface in steady state, and allow an initially
randomly arranged assembly of sliding particles to evolve on it.
In an initial short-time relaxation, particles slide down to the bottom of
local minima. After this, the density distribution evolves owing to the
rearrangement of the stochastically evolving surface, whose local slopes
guide particle motion. We found in numerical simulations that the surface
fluctuations actually drive the system towards large scale clustering
of particles. This can be seen as follows.
After time $t$, the base lengths of coarse-grained valleys
of length $t^{1/z}$ would have overturned, where $z$ is the dynamical
exponent of the surface. We thus expect that the latter length scale
sets the scale of particle clustering at time $t$. To test
this we monitored the equal time correlation function ${\cal C}(r,t)
\equiv \langle \sigma_{o}(t) \sigma_{o+r}(t) \rangle$ by Monte-Carlo
simulation. We found that it has a scaling form
\beq
C = f(r/{\cal L}(t))~~~~~~~~~~{\rm with}~~{\cal L} \sim t^{1/z}
\label{splt}
\eeq
in accord with the arguments given above. 
The data for ${\cal C}(r,t)$ for the particles on EW, KPZ and GBDT surfaces
are shown to collapse in Figs. \ref{ewsp}, \ref{kpzsp} and \ref{dtsp},
respectively. Evidently, Eq. \ref{splt} holds quite well for all three
surfaces, despite the widely different values of $z$ for the three.
The onset of scaling will be discussed further in section \ref{robust},
where we discuss the effect of varying the ratio of rates of relative
updates of the particles and the surface.

\figone{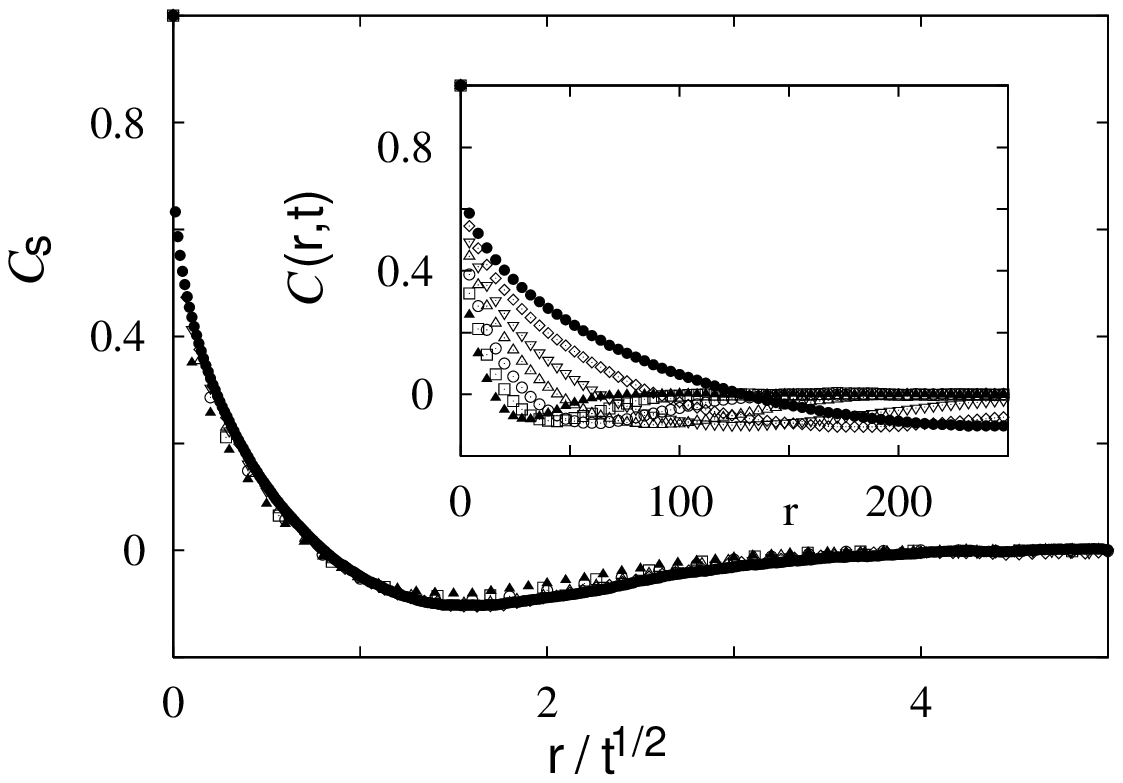}
{The data shown in the inset
for ${\cal C}(r,t)$ for the SP model with an EW surface
at different times $t = 400{\times}2^n$ (with $n = 0$,...,$6$),
is seen to collapse when scaled by ${\cal L}(t) \sim t^{1/2}$.
For small arguments, the scaling function has a cusp with $\alpha \simeq 0.5$.}{ewsp}

\figone{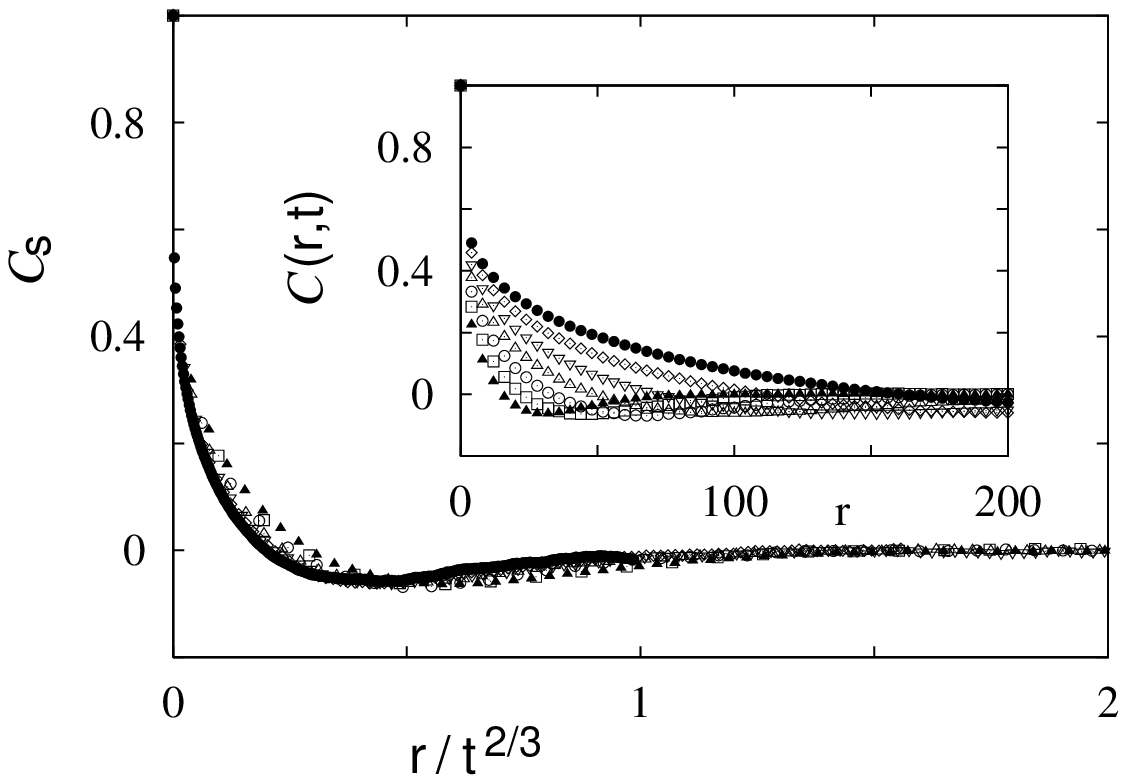}
{The data shown in the inset for ${\cal C}(r,t)$ for the SP 
model with a KPZ surface
at different times $t = 400{\times}2^n$ (with $n = 0$,...,$6$), 
is seen to collapse when scaled by ${\cal L}(t) \sim t^{2/3}$.
For small arguments, the scaling function has a cusp with $\alpha \simeq 0.25$.}{kpzsp}

\figone{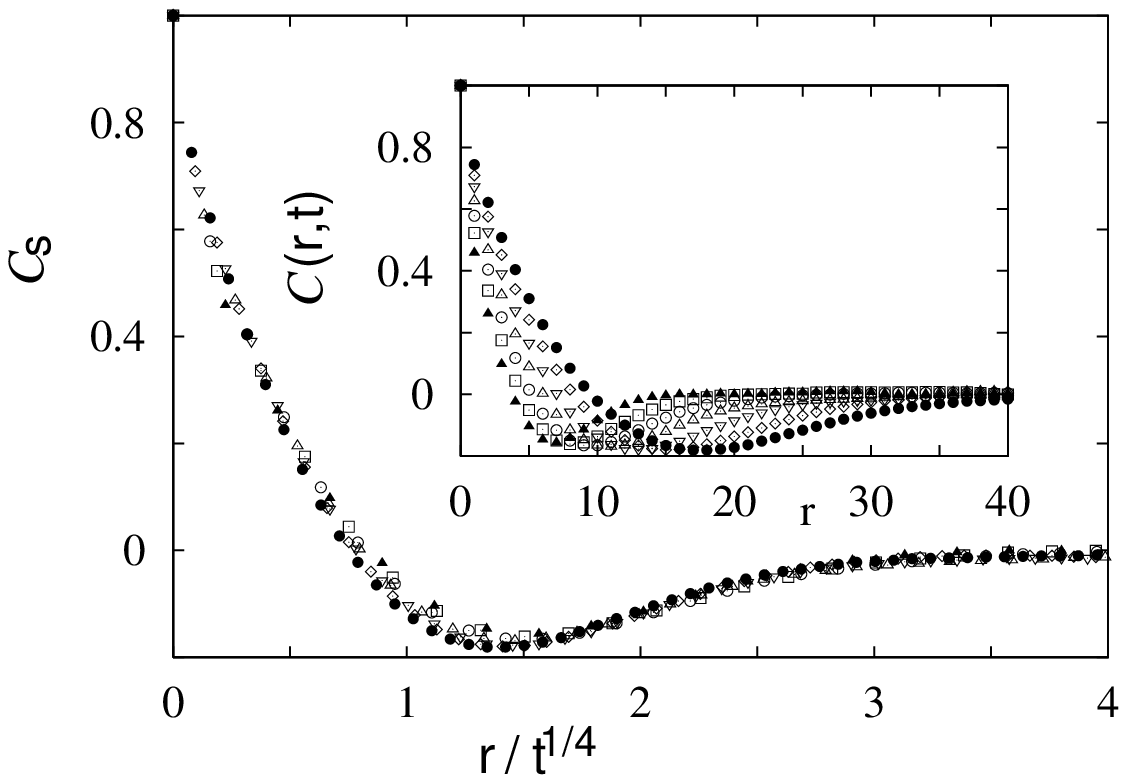}
{The data shown in the inset
for ${\cal C}(r,t)$ for the SP model with a GBDT surface
at different times $t = 400{\times}2^n$ (with $n = 0$,...,$6$),
is seen to collapse when scaled by ${\cal L}(t) \sim t^{1/4}$.
The scaling function has no cusp and  $\alpha \simeq 1.0$.}{dtsp}

To determine the short distance behavior of the decay of ${\cal C}$ as
a function of $r/{\cal L}(t)$, we evaluated the structure factor ${\cal
S}(k)$ for ${\cal C}$. For any finite ${\cal L}(t)$, we may write
\beq
{\cal C} = {\cal C}_o(r) + {\cal C}_s(r/{\cal L}),
\label{analyticcr}
\eeq
where ${\cal C}_o(r)$ is the analytic part which
decays over small distances $r$, while ${\cal C}_s$ is the nonanalytic
part which scales as a function of $r/{\cal L}$. We are primarily interested
in ${\cal C}_s$, and so need to
subtract the appropriate ${\cal C}_o$ from ${\cal C}$. In terms of the scaled
variable $y = r/{\cal L}$, ${\cal C}_o$ contributes only to $y = 0$, in the
limit ${\cal L} \rt \infty$. In that limit we write ${\cal C}(y)
= {\cal C}_s(y) + {\cal C}_o \delta_{y,0}$, and determine ${\cal C}_o$ by
seeing which value gives the longest power-law stretch for
${\cal S}/{\cal L}$, as judged by eye. In Fig.
\ref{sksub} we show ${\cal S}$ for a late time, obtained without
any subtraction and after subtraction of ${\cal C}_o \delta_{y,0}$ with
${\cal C}_s(0) = 0.71$. The power law decay as
$\sim 1/(k{\cal L})^{\alpha+1}$,
stretches over a substantially larger range in the latter case,
corresponding to a real space decay with a cusp exponent $\alpha$. A nonzero
value of $C_o$ implies that $m_c \neq 1$, as $m_c$ is given by
$\sqrt{1-{\cal C}_o}$. This indicates that the
particle-rich phase has some holes and {\it vice versa}.

\figone{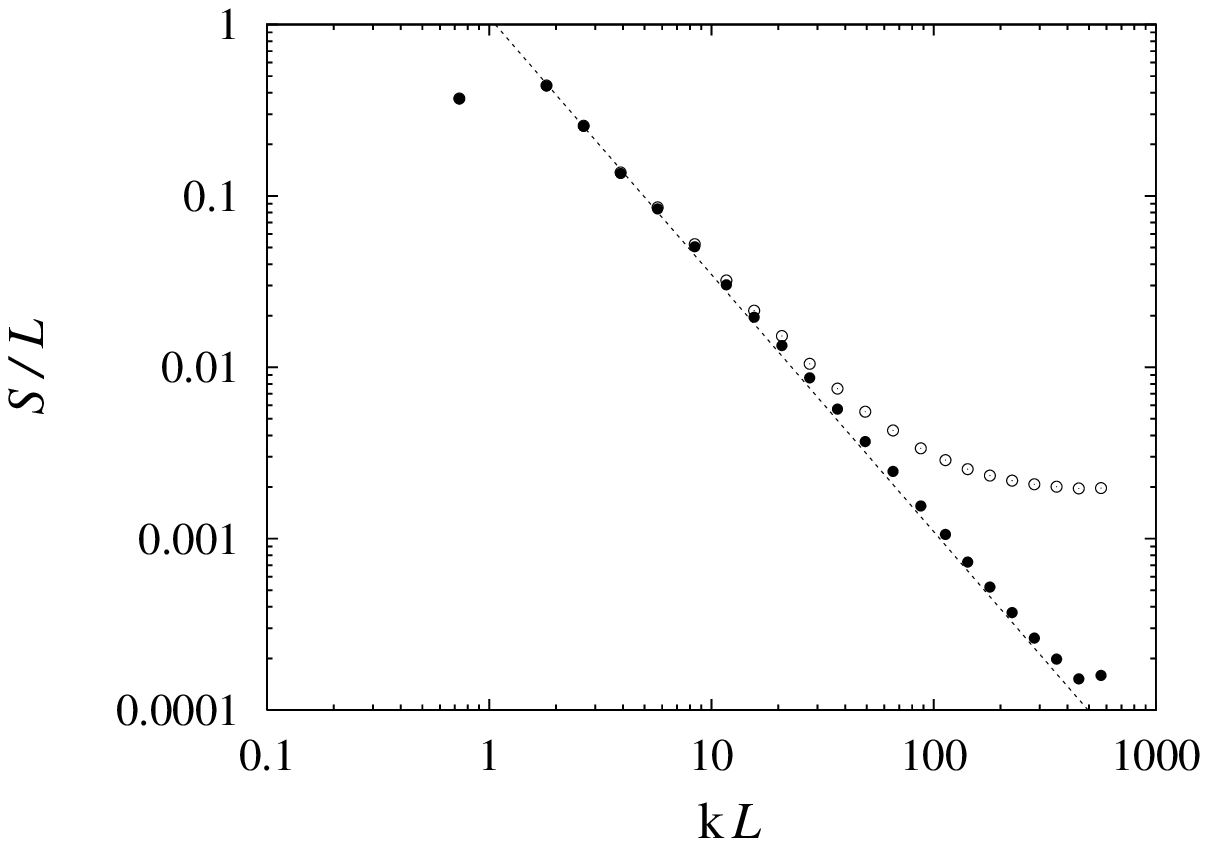}
{The scaled structure factor ${\cal S}(k)/{\cal L}$ versus
$k {\cal L}$ for the SP model with EW surface, with ($\bullet$) and
without ($\circ$) subtraction of the analytic part.}{sksub}

\figone{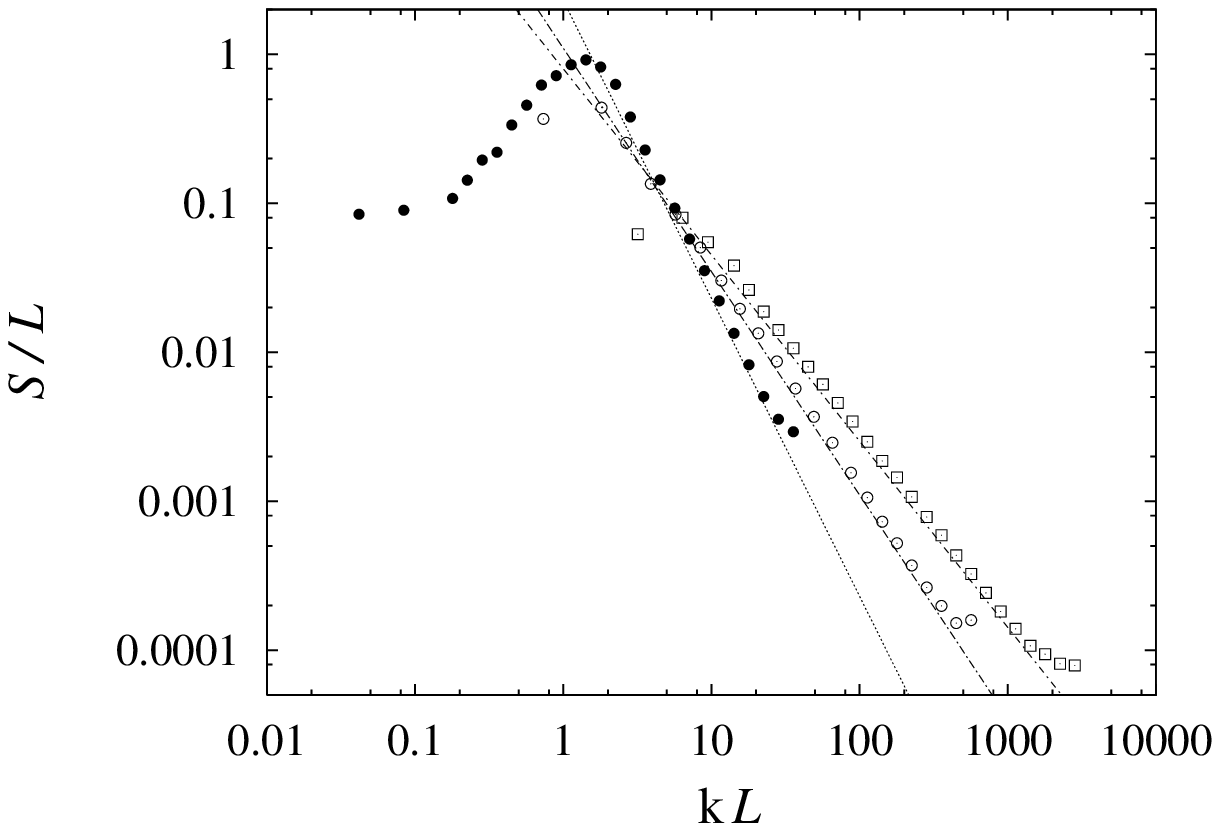}{The
scaled structure factor ${\cal S}/{\cal L}$ is plotted
against $k{\cal L}$, corresponding to the curves for $n = 6$
in the insets of Figs. \ref{ewsp}, \ref{kpzsp} and \ref{dtsp}.
The slopes at large $k{\cal L}$ for
KPZ, EW and GBDT are $-1.25$, $-1.5$ and $-2$, respectively.}{sk3}

In Fig. \ref{sk3} we show, ${\cal S}$ corresponding to the three different
surfaces at $t = 400\times 2^6$. We
find that for the EW surface $\alpha \simeq 0.5$, for the KPZ surface
it is $\simeq 0.25$, and for the GBDT surface it is $\simeq 1.0$. Thus
there is a deviation from the Porod law behavior for the EW and KPZ surface
fluctuations, and no such deviation for the GBDT surface. In all three cases,
we see that the behaviour of the 2-point functions in the particle
system resembles the corresponding correlation functions of the
CD model for the underlying surface. In the KPZ case, the value
of the exponent $\alpha \simeq 0.25$ is different from its value $\alpha = 
1/2$ in the CD model counterpart. 
For the EW and GBDT surfaces, the values of $\alpha$ are $\simeq 0.5$ and 
$1.0$ respectively as in the corresponding CD models.  


Since the SP model corresponding to the GBDT surface does not exhibit 
anomalous behaviour of the scaled two-point correlation function which
is a signature of FDPO, we 
do not consider it further in our subsequent discussion  of the 
steady state.

\subsection{Steady state of the SP model}

We first study 1-point functions in order to characterize the steady
state. As the system phase separates, a suitable quantity to study
is the magnitude of the Fourier components of the density profile
\begin{equation}
Q(k) = |{{1 \over L} {\sum_{j=1}^L} e^{ikj} n_j}|,~~~~~~~~~~
k = {2{\pi}m \over L}.
\label{qk}
\end{equation}
where $n_j=(1+\sigma_j)/2$ and
$m = 1,...,L-1$. A signature of an ordered state is that
in the thermodynamic limit, the average values $\langle Q(k) \rangle$ go
to zero for all $k$, except at $k \rt 0$. We monitored these averages for
the system of sliding particles, with the average $\langle \cdots
\rangle$ performed over the ensemble of steady state configurations.
In Figs. \ref{qkewh} and \ref{qkkpz} we show the values of $\langle Q(k)
\rangle$ as a function of $k$ for various system sizes $L$, for the
EW and KPZ surfaces respectively. In both the cases, for all $k \neq 0$
the value of $\langle Q(k) \rangle$ falls with increasing $L$ indicating
that $\langle Q(k) \rangle \rt 0$
in the thermodynamic limit, for any fixed, finite $k$. But for $k = 2\pi/L$,
we see that the value of $\langle Q(k={2\pi \over L}) \rangle$
approaches a constant. The sharpening of the curves near $k \rt 0$
implies an ordered steady state.

\figone{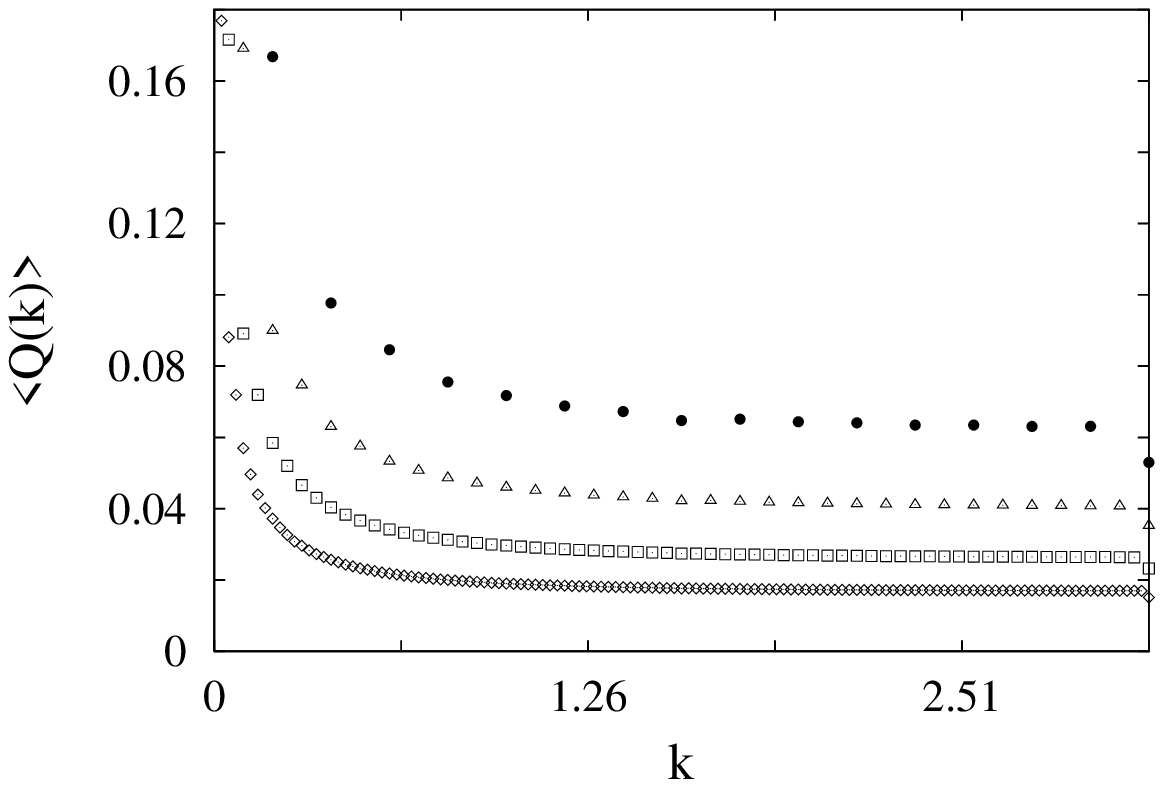}{$\langle Q(k) \rangle$ plotted as
function of $k = 2\pi m/L$,
for different system sizes $L = 32$, $64$, $128$ and $256$, for an EW surface.}
{qkewh}

\figone{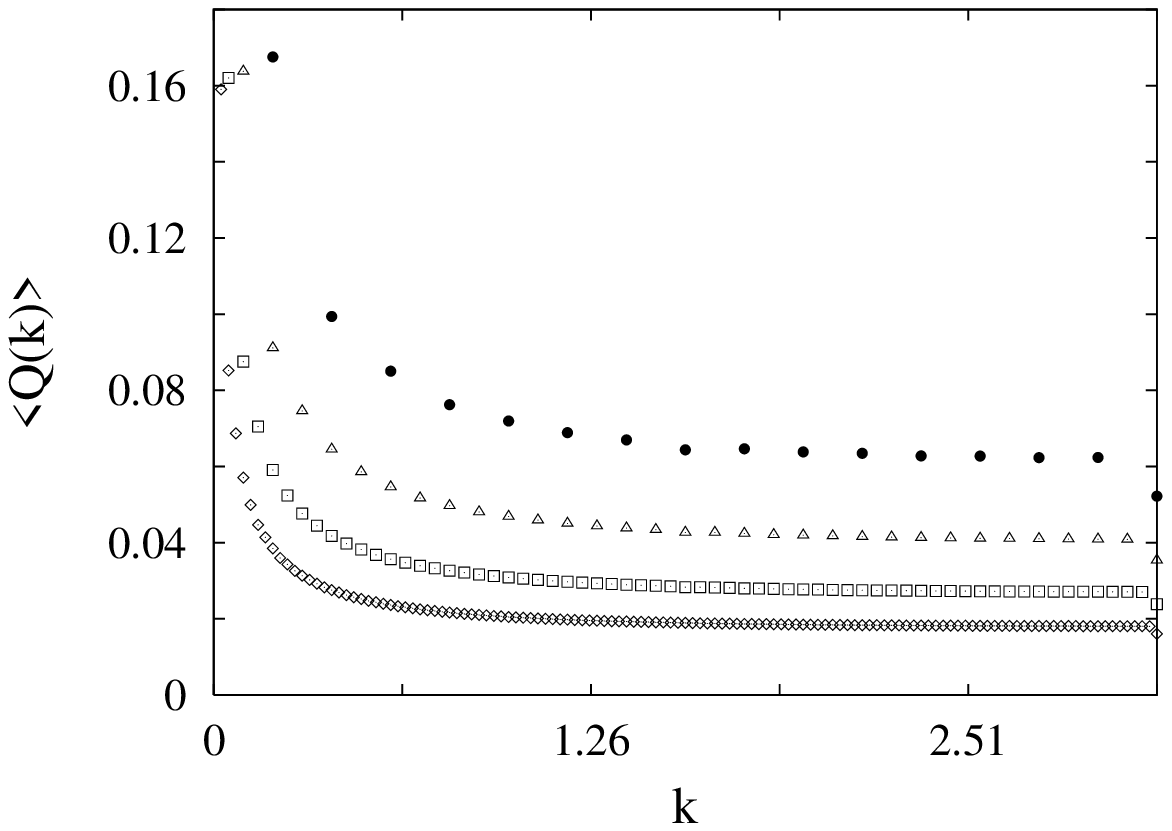}{$\langle Q(k) \rangle$
plotted as function of $k = 2\pi m/L$
for different system sizes $L = 32$, $64$, $128$ and $256$, for a KPZ
surface.}{qkkpz}

The above behavior of $\langle Q(k) \rangle$ as a function of $k$
suggests that we take the value $Q^* \equiv Q({2\pi \over L})$,
(corresponding to $m=1$) as a measure of the extent
of phase separation. We have used $Q_1 = \langle Q^* \rangle$ as the order
parameter earlier also for the CD3 model, and note that it
has been used earlier in other
studies of phase-separated systems \cite{Korniss5}. Here we find
that $Q_1 \simeq 0.18$ and $0.16$ for particles on the
EW and and KPZ surfaces respectively. The latter values being nonzero
indicates that the steady state is ordered. At the same time, the values
being less than $0.318$ indicates that the states deviates substantially from
a phase separated state with two completely ordered domains. To have a full
characterization of the fluctuations which dominate the ordered
state, one should actually evaluate the probability distributions of
all the $Q$'s, e.g. $Q^* = Q({2\pi \over L})$, $Q(2) = Q({4\pi \over L})$,
$Q(3) = Q({6\pi \over L})$, $\cdots$. We show below (in Fig. \ref{probq1})
one of these distributions, namely that of
$Q^* = Q({2\pi \over L})$, for an EW surface.
We find that the distribution $P(Q^*)$ remain broad (with root-mean-square
deviation being $\simeq 0.07$) even as $L \rt \infty$,
indicating again the dominance of large scale fluctuations.

\figone{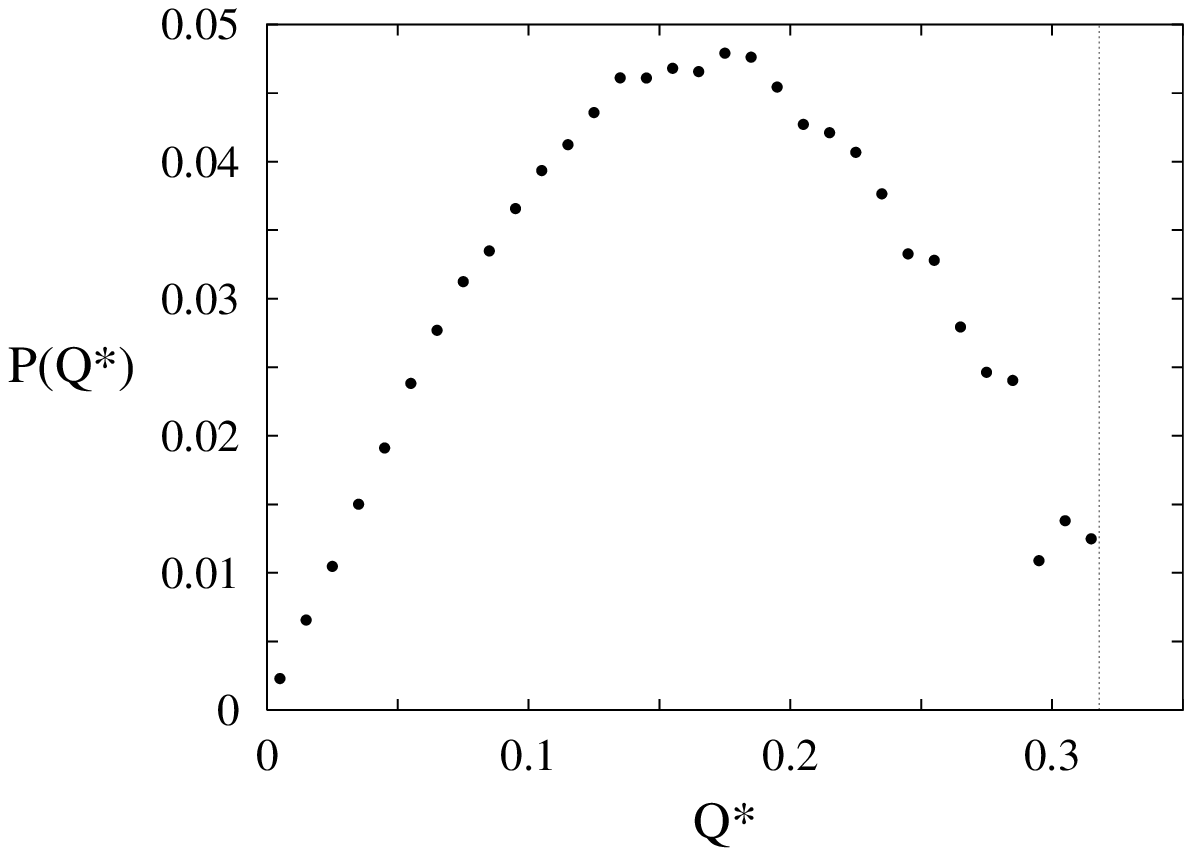}{Numerically
determined probability distribution $P(Q^*)$
of the order parameter $Q^*$, obtained for the SP model with EW surface
fluctuations, in steady state.}{probq1}

It is instructive to monitor
the variation of $Q^*$ as a function of
time $t$, for different system sizes. For an EW surface (Fig. \ref{q1tL})
the value of $Q^*$ shows strong excursions
about its average value, consistent with the broad distribution shown in
Fig. \ref{probq1}. The temporal separation period of these fluctuations
of the order parameter increases roughly as $\sim L^2$, but their amplitude is
independent of $L$. Consequently $P(Q^*)$ approaches an $L$-independent
form as $L \rt \infty$. A temporally oscillatory order parameter also has
been found earlier in a model for comparative learning \cite{Luck5}. However
the temporal behaviour in our case is quite different from the almost periodic
fluctuation in the latter model, as the Fourier spectrum of the
time series in $Q^*(t)$ in our case follows a broad power-law.
We have not pursued a detailed study of the temporal behaviour any
further.

\figone{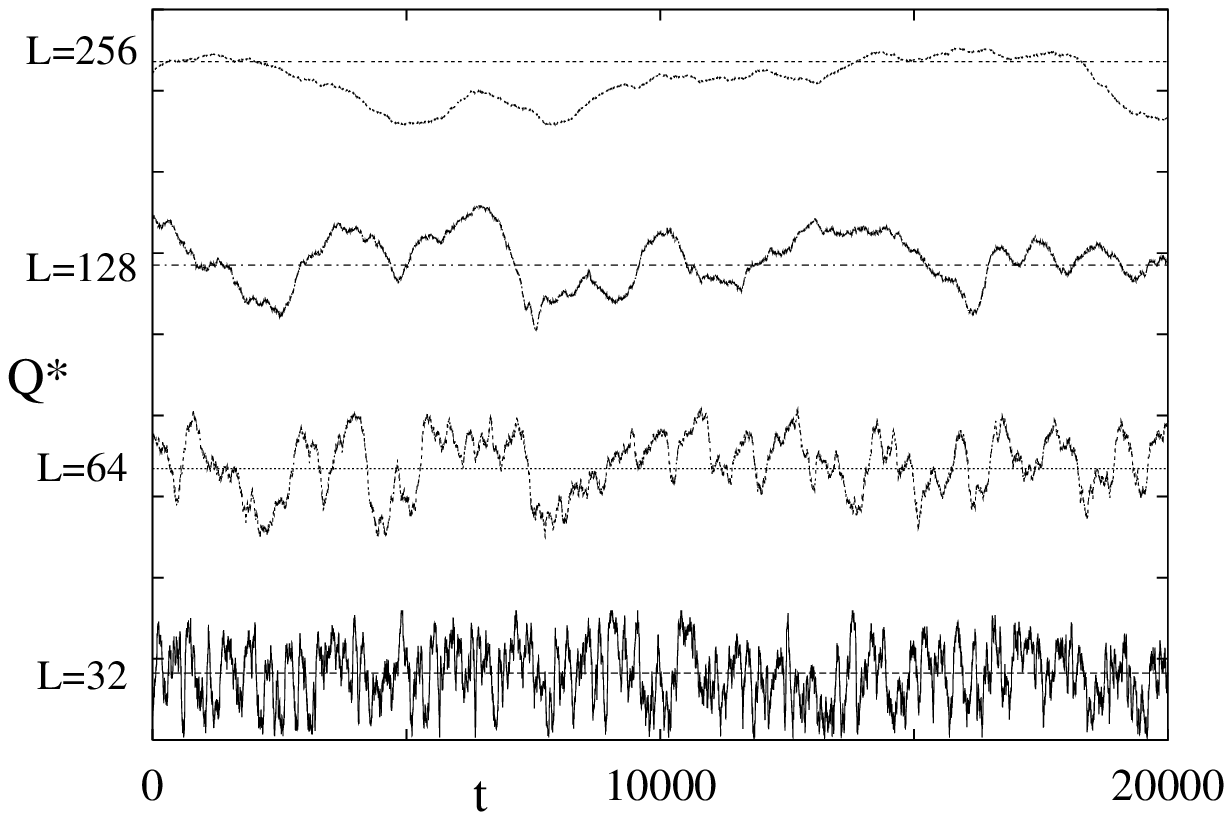}{Variation
of $Q^*$ with time $t$, for different system
sizes $L = 32$, $64$, $128$ and $256$, showing that the separation between the
fluctuations of the order parameter increases with $L$, but that the
amplitude does not vary much.}{q1tL}

\figone{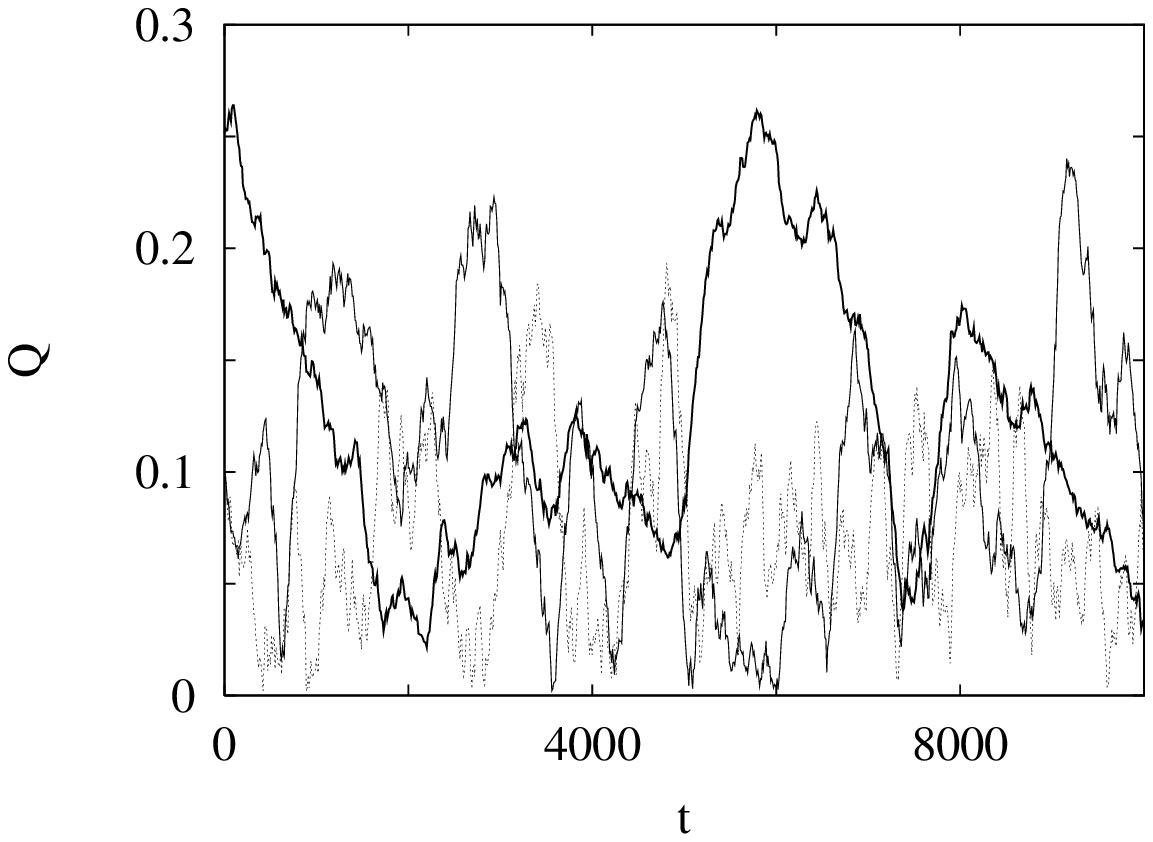}{Variation
of $Q^* \equiv Q(1)$ (solid thick line), $Q(2)$ (solid thin line) and
$Q(3)$ (broken line) are shown as a
function of time to show that a decrease in value of one is accompanied
by an increase in the others, indicating that one large cluster may break up
into a few large ones, in the steady state. The system size 
was $L = 128$.}{q123t}

The fluctuation of $Q^*$ in Fig. \ref{q1tL}, gives rise to an interesting
question: Does the system become disordered and lose the phase ordering
property when the value $Q^*$ falls
to low values? The answer is no, as is very clearly brought out in
Fig. \ref{q123t} in which $Q^* \equiv Q(1)$, $Q(2)$,
$Q(3)$ have been plotted simultaneously as a function of
time $t$ for a single evolution of the system. We observe that a dip
in $Q^*$ is accompanied by a simultaneous rise in the value of either
$Q(2)$ or $Q(3)$. This implies that whenever the system loses a
single large cluster (making $Q^*$ small) either two or three such
clusters appear in its place (making the values of $Q(2)$ and $Q(3)$ go up).
Thus the system remains far from the disordered state, and always has a
few large particle clusters which are of macroscopic size $\sim L$.
A numerical
study showed that the average size of the largest particle cluster 
$\sim 0.14 L$.

We have seen above that in the SP models, the order parameter has a broad
distribution just as in their CD model counterparts. We observe further
that the particle and hole cluster size distributions in the steady state
of the SP model decay as a power-law: $P(l) \sim l^{-{\theta}}$. In Fig.
\ref{plcrspew} for the EW surface,
we find that the particle (denoted by symbols) and hole
(denoted by lines) distributions coincide, with $\theta \simeq 1.8$. By
contrast, Fig. \ref{plcrspkpz} for the KPZ surface
shows that the particle and hole distributions
are not identical. This is because with asymmetric rates
($p_1 \neq q_1$), the surface has an overall motion in one direction,
such that the downward motion of the particles and the upward motion
of the holes, due to gravity, are no longer symmetrical.
We checked that the distributions for particles
and holes get interchanged if the rates $p_1$ and $q_1$ are interchanged.
The exponent for the decay of both the particle and hole distributions is
$\theta \simeq 1.85$.

\figone{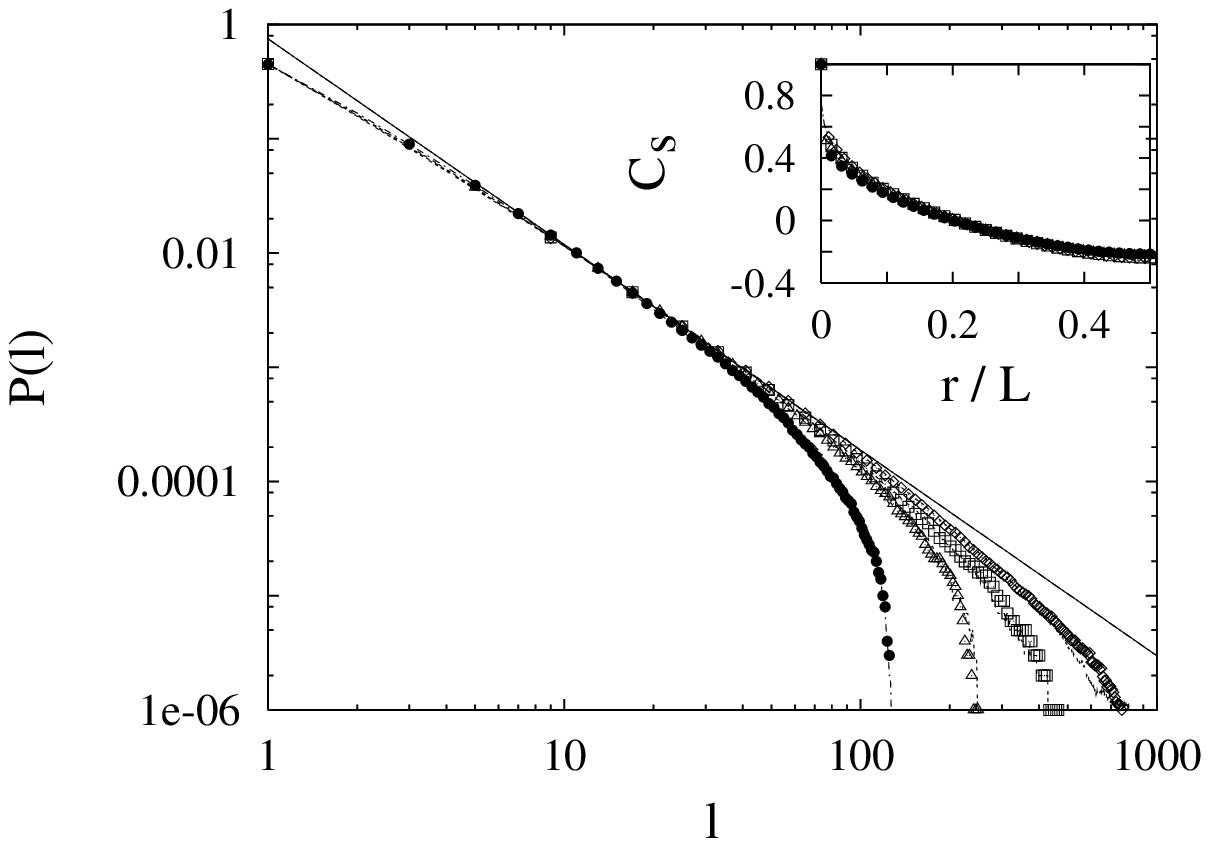}{$P(l)$ vs. $l$
for clusters of particles (symbols)
and holes (lines) in the SP model with an EW surface, for
different system sizes $L=256$, $512$, $1024$ and $2048$. $P(l)$ decays as a
power law with $\theta \simeq 1.8$. The inset shows collapsed data of
steady state $C(r)$ for $L=64$, $128$, $256$ and $512$ as a function
of $r/L$; the scaling function has a cusp with
$\alpha \simeq 0.5$.}{plcrspew}

\figone{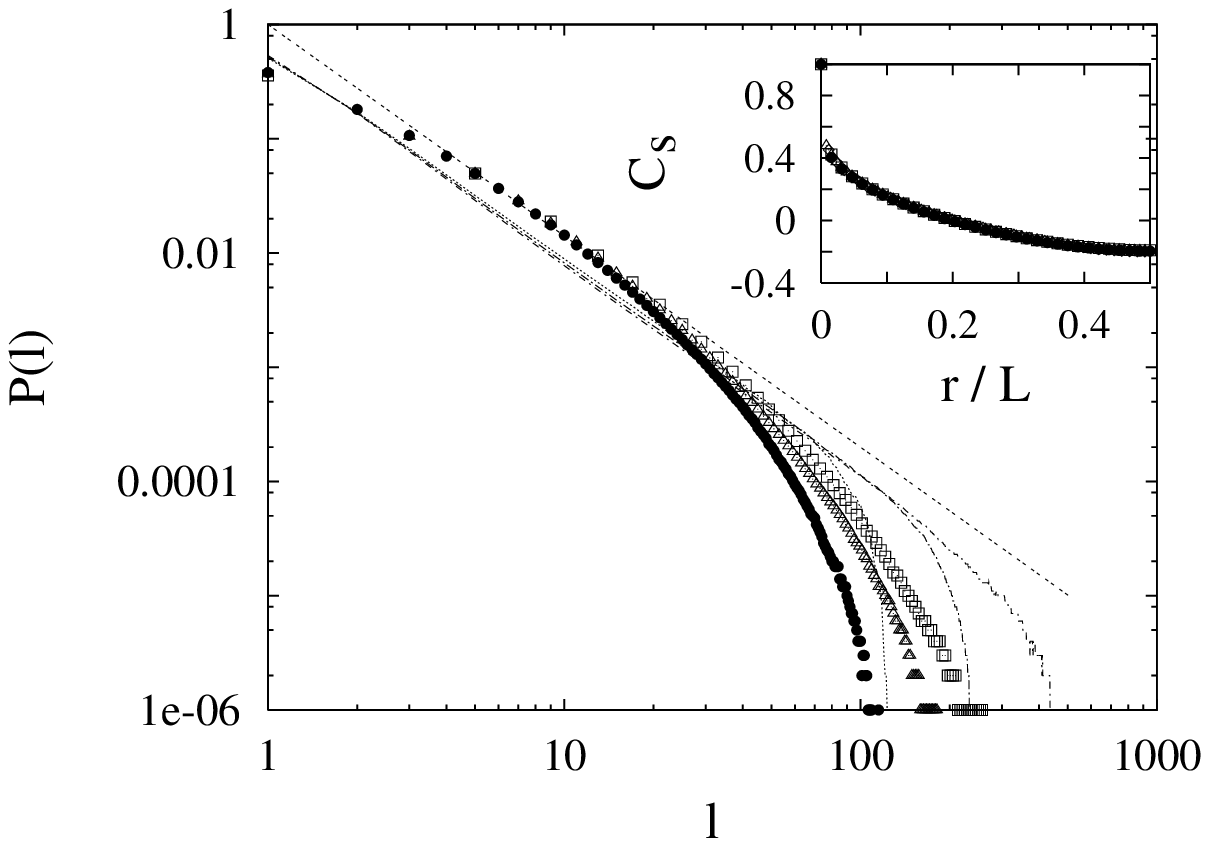}{$P(l)$ Vs. $l$
for clusters of particles (symbols)
and holes (lines) in the SP model  tha KPZ surface, for
different system sizes $L=256$, $512$, $1024$ and $2048$. The data shows 
the existence of a
particle-hole asymmetry. A power-law with $\theta \simeq 1.85$ has been shown
along with the curves as a guide to the eye. The inset shows collapsed data of
steady state $C(r)$ for $L=64$, $128$, $256$ and $512$ as a function
of $r/L$; the scaling function has a cusp with
$\alpha \simeq 0.25$.}{plcrspkpz}

Finally we note that the 2-point correlation functions in the steady
state of the SP model exhibit a scaling form in $r/L$ and have the
same cusp exponents as in the coarsening regime (with $\cal L$ being
replaced by $L$). For the EW surface,
the scaling curve shown in the inset of Fig. \ref{plcrspew} exhibits a
cusp with $\alpha \simeq 0.5$. The corresponding curve for
the KPZ surface, shown in the
inset of Fig. \ref{plcrspkpz} also exhibits a cusp, with $\alpha
\simeq 0.25$.  The fact that $m_c < 1$ in these curves, as for those in
the coarsening regime of the SP model, is indicative of the fact that
the pure phase which are particle rich also have holes in them. In
this respect, the pure phases differ from their CD model counterparts.

We have seen above that the FDPO of the sliding particles in the
SP model is qualitatively of the same type as in the CD models for the
underlying surfaces. We measured the average overlap $O = \langle
s_i \sigma_i \rangle$ to get a quantitative estimate of the extent of
correlation between the sliding particles (holes) and the valleys
(hills) of the underlying surface. We found that it is nonzero as
we expected, e.g. for the EW surface $O \simeq 0.26$ and $0.39$
corresponding to $s_i$ being defined within CD2 and CD3 models. The
overlap is greater in case of CD3 model, since the domains are most often
smaller than $L/2$ and this matches with the fact that particles clusters
are also of size $\leq L/2$. On the other
hand, domains in the CD2 model can be almost as big as $L$. For the
KPZ surface, $O \simeq 0.26$ corresponding to the overlap
between particles and the coarse grained depth variables 
$\{s_i\}$'s of the CD3 model.

\section{Robustness of FDPO}
\label{robust}

We did several numerical tests to check the robustness of the fluctuation
dominated ordered state for the sliding particle (SP) problem.

\noindent (i) We explored
the effect of varying the ratio $R = p_2/p_1$, the relative rate
at which the particles get updated as compared to the surface.

\noindent (ii) We allowed the possibility of a small but finite
rate ($q_2 \neq 0$) of the particles to hop uphill on a local $\tau$ slope.

\noindent (iii) We made the overall slope nonzero, in the case
of the KPZ surface.

We found that FDPO stays with (i) and (ii), while it is lost with (iii).

For the EW surface, with $R = 0.2$ (i.e.  the surface moving $5$ times slower
than the sliding particles), we found that $Q_1$
remains close to but slightly larger than
$0.18$, the value for $R = 1$. We checked the
correlation function ${\cal C}(r,t)$ in the coarsening regime, and
found that it has a cusp as a function of $r/{\cal L}$ with the
exponent $\alpha \simeq 0.5$. For $R = 5$ (i.e. the surface moving $5$ times
faster), we found $Q_1 \simeq 0.15$. The latter value indicates
a lesser degree of ordering and this is also mirrored in the 2-point
function ${\cal C}(r,t)$: the collapse of the data as a function of
$r/{\cal L}$ occurs beyond a time which is greater than that for $R =
1$, i.e. the scaling regime sets in much later. Nevertheless, at large
enough times, the cusp exponent is unchanged ($\alpha \simeq
0.5$). Figure \ref{skrates} (lower curves)
shows the log-log plot of ${\cal S}/{\cal L}$ versus $k{\cal L}$
for the three rates $R = 5$, $1$ and $0.2$.
All of them have slopes $-1.5$, which indicate $\alpha \simeq 0.5$.

\figone{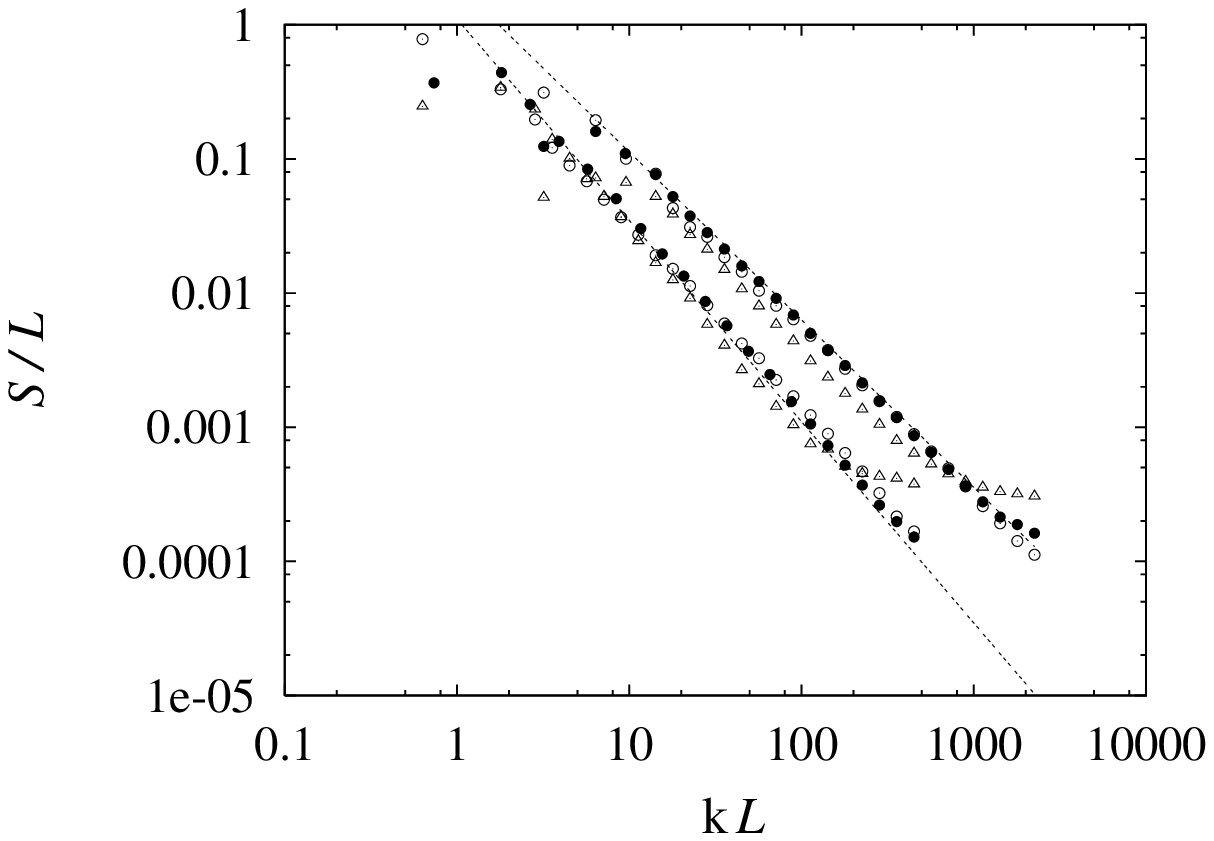}
{${\cal S}/{\cal L}$ is plotted against
$k{\cal L}$, with $R = 5$ (triangles), $R=1$ ($\bullet$) and
$R=0.2$ ($\circ$) for the EW and KPZ surfaces at $t = 400 \times 2^6$.
For clarity of display, we have multiplied the data for the KPZ
surface by a factor of $2$.}{skrates}

A similar evaluation of ${\cal C}$ was also done for the KPZ
surface, and is also shown in Fig. \ref{skrates} (upper curves).
The observed slope of $-1.25$ implies that the
cusp exponent remains $\alpha \simeq 0.25$ for all of them.

We conclude that the variation of update rates affects the degree of ordering
but not the asymptotic scaling properties, as indicated by the constancy
of the cusp exponent $\alpha$.

\figone{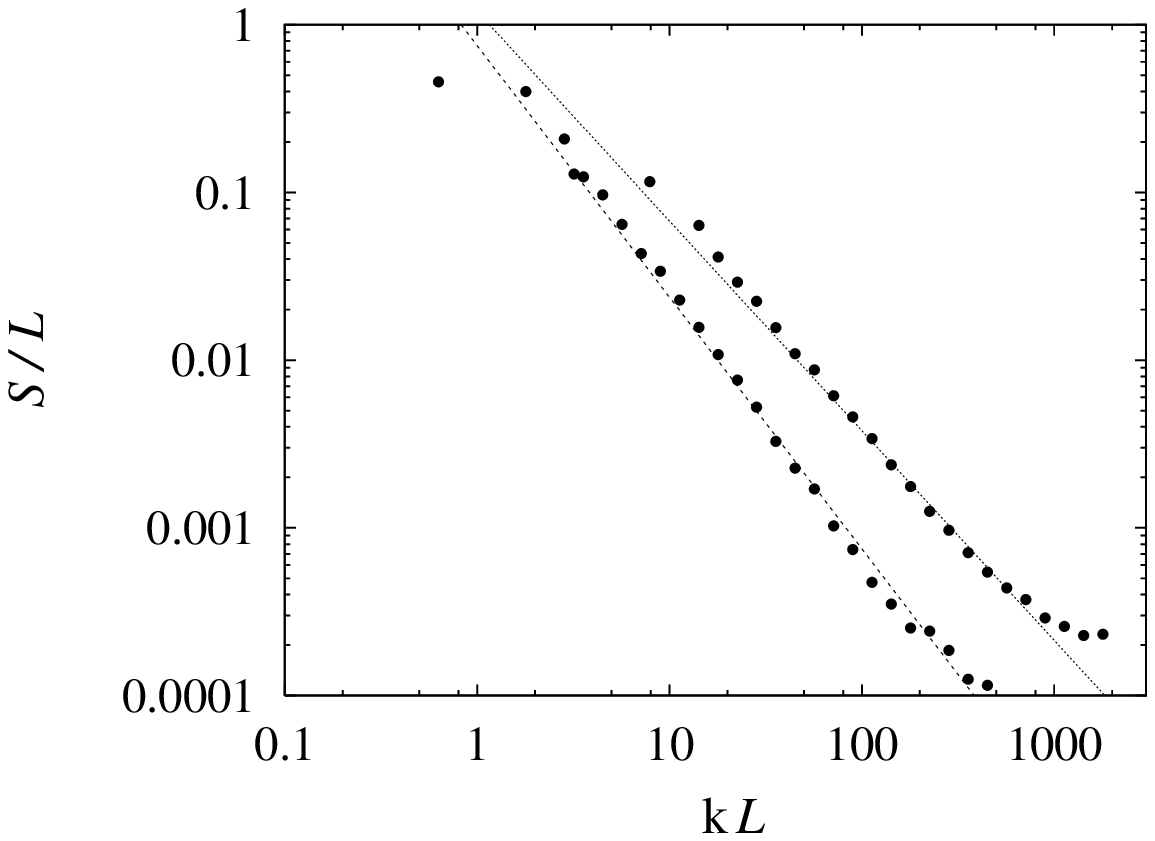}
{${\cal S}/{\cal L}$ is plotted against
$k{\cal L}$, for $t = 400 \times 2^6$, with a finite uphill
hopping rate for both EW and KPZ surfaces. The data for the KPZ
surface is multiplied by a factor of $2$ for clarity of display.}{sk_upneq0}

So far we have considered uphill hopping rate to be strictly zero, i.e. $p_2/q_2 =
\infty$. By allowing for $q_2 \neq 0$, i.e. allowing for upward motion of
particles, we saw that the FDPO persists, so long as $p_2 > q_2$. In
Fig. \ref{sk_upneq0}, we show ${\cal
S}(k)/{\cal L}$ as a function of $k {\cal L}$, at a large time $t$ for
EW and KPZ surfaces respectively, when the ratio $p_2/q_2 = 5$. We
find that the slopes are $-1.5$ and $-1.25$ in the two cases
indicating that the values of the cusp exponents are still $\alpha
\simeq 0.5$ and $\alpha \simeq 0.25$ respectively, for the two surfaces.

This points to the universality of the value $\alpha
\simeq 0.5$ (EW) and $\alpha \simeq 0.25$ (KPZ) over a range of models with
different values of $R$, and also with respect to varying $p_2/q_2$.

\figone{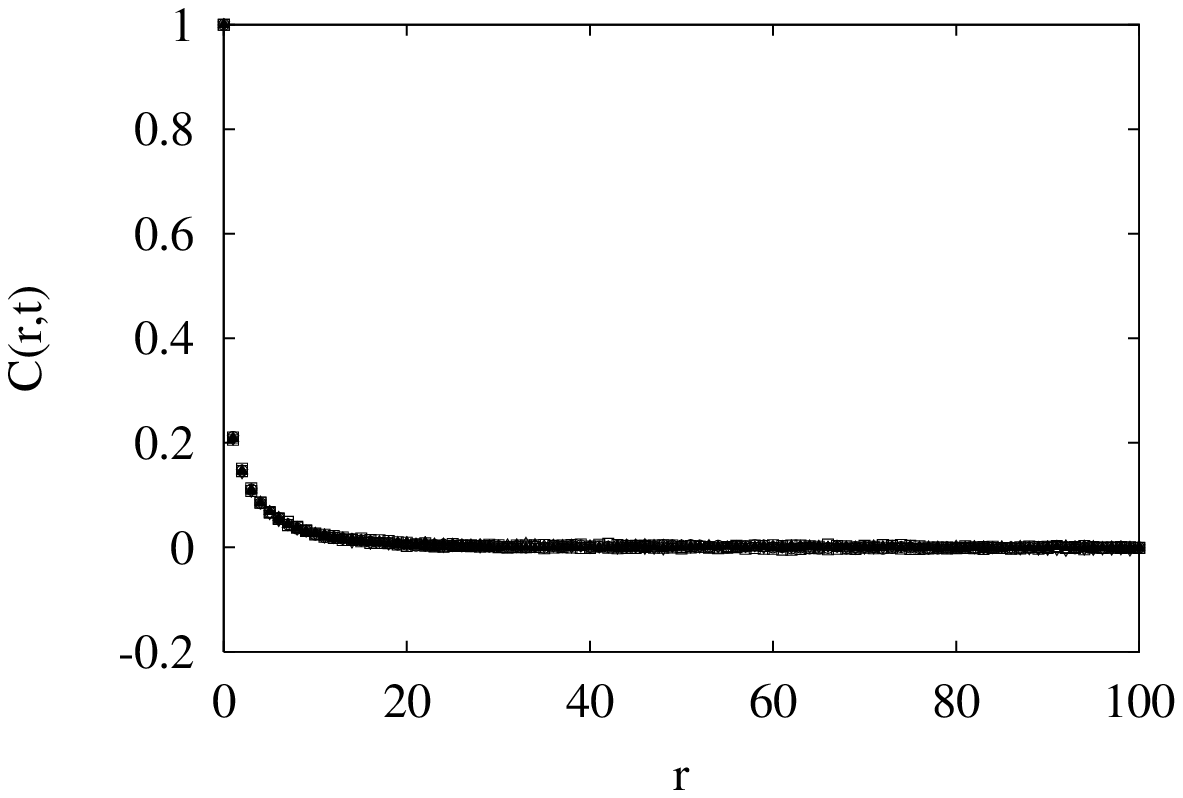}{For a tilted KPZ surface,
the curves for ${\cal C}(r,t)$ as a function of $r$ all overlap
at different times $t = 400 \times 2^n$ (with $n = 0,...,6$),
indicating that there is no growing length scale ${\cal L}(t)$.
Thus tilt removes FDPO.}{kinekpzspcr}

We also investigated the effect of having an overall tilt of the KPZ
surface.  This leads to an overall movement of the transverse surface
fluctuations, which are the analogues of
kinematic waves in particle systems \cite{lighthill5,kinematic5}. In
the presence of such a wave, the profile of hills and valleys of the
surface sweep across the system at finite speed, and the particles do
not get enough time to cluster.  Consequently the phenomenon of FDPO is
completely destroyed. In Fig.  \ref{kinekpzspcr} we show ${\cal
C}(r,t)$ as a function of $r$ (there is no scaling by ${\cal L}(t)$)
for several $t$. The curves are
independent of $t$, in the absence of coarsening towards a phase ordered 
state. 

\section{Conclusion}
\label{Conclusion}

In this paper we have discussed the possibility of phase ordering of a sort
which is dominated by strong fluctuations. In steady state, these
fluctuations lead to variations of the order parameter of order unity, but
the system stays ordered in the sense that with probability one, a finite
fraction of the system is occupied by a single phase. The value of this fraction
fluctuates in time, leading to a broad probability distribution of the order
parameter.

We demonstrated these features in two types of models having to do with
surface fluctuations --- the first, a coarse-grained depth (CD) model
where we could establish these properties analytically, and the second a
model of sliding particles (SP) on the surface in question. For these models
we found that besides (a) the broad probability distribution of
the order parameter  (which we may take to be the defining characteristic
of FDPO), the steady state was also characterised by (b) power laws of
cluster size distributions and (c) cusps in the scaled two-point
correlation function, associated with the breakdown of the Porod law. The
connection between (b) and (c) was elucidated using the independent interval
approximation. Further, an extremal statistics argument showed that the
largest cluster drawn from the power law distribution is of the order of the
system size; this implies a macroscopic ordered region,
so that within our models, properties (a) and (b) are connected.

There are several open questions.
Does fluctuation-dominated phase ordering occur in other,
completely different types of systems as well ? Are properties (b) and (c)
necessarily concomitant with the defining property (a) of FDPO?
Can one characterize quantitatively the dynamical behaviour in
the FDPO steady state?

Our model of particles sliding on a fluctuating surface relates
to several physical systems of interest. First, it describes a new
mechanism of large scale clustering in vibrated granular media, 
provided the vibrations are random both in space and time. Second,
it describes a special case (the passive scalar limit) of a crystal
driven through a dissipative medium, for instance a sedimenting
colloidal crystal \cite{4authors}. Finally, related models describe
the formation of domains in growing binary films \cite{Drossel5}.
It would be interesting to see if ideas related to FDPO play a role
in any of these systems.

It would also be interesting to examine fluctuating phase-ordered states
in other nonequilibrium systems from the point of view of FDPO.
For instance, in a study of jamming in the 
bus-route model studied in \cite{oloan}, the largest
empty stretch in front of a bus was 
found to be of order $L$, and it is argued that
such a stretch survives for a time which is proportional to $L^2$ for a 
nonvanishing rate of arrival of the passengers. These
features are reminescent of the behaviour of the CD and SP models derived
from the Edwards-Wilkinson model discussed above. 
However, more work is required to make a clear statement about FDPO in
the bus-route model.

In general,
fluctuation-dominated phase ordering is evidently a possibililty that
should be kept in mind when discussing new situations involving phase
ordering in nonequilibrium systems, both in theory and in experiment.

We acknowledge useful discussions with M.R. Evans, D. Dhar, G. Manoj,
S. Ramaswamy and C. Dasgupta.

\end{multicols}

\end{document}